**Classification:**

PHYSICAL SCIENCES; Chemistry

# *"On demand" triggered crystallization of CaCO$_3$ from solute precursor species*

Tomasz M. Stawski[1,2], Teresa Roncal-Herrero[3,4], Alejandro Fernandez-Martinez[5], Adriana Matamoros Veloza[2], Roland Kröger[3], Liane G. Benning[1,2,6]

[1]German Research Centre for Geosciences, GFZ, Interface Geochemistry, Potsdam, Germany; [2]School of Earth and Environment, University of Leeds, Leeds, UK; [3]Department of Physics, University of York, York, UK; [4] Institute of Microwaves and Photonics, University of Leeds, Leeds, UK; [5]Université Grenoble Alpes, CNRS, ISTerre, Grenoble, France; [6]Department of Earth Sciences, Freie Universität Berlin, Berlin, Germany.

**Corresponding authors**:

1. Tomasz M. Stawski; German Research Centre for Geosciences, GFZ, Interface Geochemistry, 14473, Potsdam, Germany; +4933128827511; stawski@gfz-potsdam.de
2. Roland Kröger; Department of Physics, University of York, YO10 5DD, York, UK; +441904324622; roland.kroger@york.ac.uk
3. Liane G. Benning; German Research Centre for Geosciences, GFZ, Interface Geochemistry, 14473, Potsdam, Germany; +4933128828970; benning@gfz-potsdam.de

**Abstract:** Can we control the crystallization of solid CaCO$_3$ from supersaturated aqueous solutions and thus mimic a natural process predicted to occur in living organisms that produce biominerals? Here we show how we achieved this by confining the reaction between Ca$^{2+}$ and CO$_3^{2-}$ ions to the environment of nanosized water cores of water-in-oil microemulsions. Using a combination of *in situ* small-angle X-ray scattering, high-energy X-ray diffraction, and low-dose liquid-cell scanning transmission electron microscopy, we elucidate how the presence of micellar interfaces leads to the formation of a solute $CaCO_3$ phase that can be stabilized for extended periods of time inside micellar water nano-droplets. We can also control and "on-demand" trigger the actual precipitation and crystallization of solid $CaCO_3$ phases through the targeted removal of the organic-inorganic interfaces.

**Statement of Significance:** Mixing two solutions, one containing calcium ions ($Ca^{2+}$) and one carbonate ions ($CO_3^{2-}$) at high concentrations, leads to the instantaneous and uncontrollable precipitation of solid calcium carbonate ($CaCO_3$). Inside e.g. marine single celled algae, this reaction is not just meticulously triggered but also highly controlled. How to reproduce this process was so far not understood. Here, we report how by using a water-in-oil microemulsion, we managed to stabilize a mixture of $Ca^{2+}$ and $CO_3^{2-}$ ions as a solute phase, and prevent its precipitation to solid $CaCO_3$. Nevertheless, we can also trigger the formation of solid $CaCO_3$ "on demand" and demonstrate that the organic-inorganic interfaces in microemulsion nanodroplets are a plausible analogue for $CaCO_3$ precipitation in living organisms.

\body

Calcium carbonate (CaCO$_3$) is the most ubiquitous functional biomineral in nature. Its mineralization by microorganisms underpins a vast array of Earth system processes, and controls the global carbon cycle.(1) In living organisms metabolic processes regulate the self-assembly pathways and morphologies of crystalline CaCO$_3$ phases(2–4), and key aspects of such CaCO$_3$ (bio)mineralization reactions have been experimentally reproduced through the use of organic-inorganic interfaces(5–10). In particular, reverse micellar systems (water-in-oil microemulsions) have attracted considerable attention in biomimetic studies since they provide a well-defined and tunable organic-inorganic interface confining the reaction environments to water droplets of just a few nanometers (hence the notion of *nanoreactors*(7, 11)). Because of this confinement and the presence of a surfactant interface, the water molecules within these nano-droplets exhibit slower dynamics in comparison to bulk water(12, 13). Therefore, such water-in-oil microemulsions represent plausible cell membrane analogs(14, 15). Thus they allow for the study of biomineral precipitation reactions in which the formation pathways are controlled by the interface and very changes in the physical matrix of the confined reaction medium (e.g. water structure), and not by the chemical reactions with the functional organic molecules. Experimentally, this is realized by mixing and diffusive exchange of ions between two separate microemulsions containing distinct dissolved salt ions(7, 14, 16–19). For the CaCO$_3$ system, several studies reported that crystalline particles could be obtained and morphologically controlled by mixing microemulsions containing aqueous solutions of CO$_3^{2-}$ and Ca$^{2+}$ at concentrations highly supersaturated with respect to any solid CaCO$_3$ polymorph(20–24). In all these studies the underlying assumption was that all observed solid CaCO$_3$ phases and morphologies precipitated directly from the mixed microemulsions, as it was deduced from *ex situ* imaging and analysis of the final reaction products.

However, reverse micelles are inherently dynamic and sensitive to changes in physicochemical properties and solely *ex situ* studies or analysis of final products are insufficient to fully quantify a reaction mechanism at such dynamic organic-inorganic interfaces(14). Therefore, in the present study we demonstrate how through multi-length-scale *in situ* and time-resolved analyses of bioinspired CaCO$_3$ formed from reverse microemulsions, we gain mechanistic insights into the CaCO$_3$ mineral growth pathways. Using a combination of *in situ* small-angle X-ray scattering (SAXS), high-energy X-ray diffraction (HEXD), and low-dose liquid-cell scanning transmission electron microscopy (LC-STEM), we elucidate how the presence of a micellar interface leads to the formation of a solute $CaCO_3$ phase that is stabilized for extended periods of time inside micellar aqueous nano-droplets. The nucleation and growth of any solid $CaCO_3$ polymorph from such nano-droplets is prevented despite the fact that the water cores in the used microemulsion were highly supersaturated with respect to all known calcium carbonate solid phases. The actual precipitation and crystallization of solid $CaCO_3$ could be triggered "on-demand" through the targeted removal of the organic-inorganic interface. Our data clearly demonstrate that the confinement within a micellar-like environment inside living organisms could be the key factor controlling mineral precipitation at high supersaturation levels.

## Results

*Mixing and diffusion between two microemulsions*

We started our experiments with two initially clear reverse microemulsions containing 0.15 mol/L $CaCl_2$ and 0.15 mol/L $Na_2CO_3$ solutions in the water cores. Analyzing these initial non-mixed single salt reverse micelles by SAXS revealed water cores with an average diameter of ~3.4 nm. Including the surfactant film this resulted in a total diameter of the reverse micelles of ~7 nm (see Note S1, Fig. S1, Table S1). Furthermore, our SAXS data confirmed that the chemistry of the dissolved salts in the aqueous cores, did not affect the size or shape of the micellar water pools (Fig. S1). After characterizing the individual microemulsions we mixed equal volumes of the two initially clear $Ca^{2+}$ and $CO_3^{2-}$ containing microemulsions, resulting in an equivalent of 0.075 mol/L $CaCO_3$ in the aqueous sub-phase. Based on previous work(18, 20, 21), this "bulk" mixed solution is predicted to exchange $Ca^{2+}$ and $CO_3^{2-}$ ions between the water cores of the two micelles, and thus create an aqueous phase supersaturated with respect to all solid $CaCO_3$ phases (see Methods for the respective saturation indices). We followed and quantified the evolution of such mixed systems through *in situ* and time-resolved SAXS experiments for up to 180 minutes. Our results (Fig. 1) revealed that at high scattering vector magnitudes, $q > 0.2$ nm$^{-1}$, the SAXS patterns were dominated by scattering from the core-shell droplets of the reverse micelles and that this high-*q* region was practically identical to scattering patterns of the non-mixed single salt micelles (see Fig. S1). Analyzing the change in scattering in this high-*q* region and utilizing the $I(q)q^4$ *vs*. *q* representation (inset in Fig. 1) allowed us to show that the average volumes of the water cores of the individual micelles decreased by approx. 5%, indicating only a minor change in the chemical environment inside the micelles. In contrast, a major (~ 500%) increase observed in the scattering intensity at $q < 0.2$ nm$^{-1}$ during the first 90 min, indicated attractive interactions between the individual micelles(25–27), and this pointed to the gradual development of larger scattering objects. After ~90 min no further change in the system was observed indicating that these large objects became relatively stable. The dramatic low-$q$ increase in intensity and the fact that at $q > 0.2$ nm$^{-1}$ the scattering curves from the freshly mixed (e.g., 3 min) and the aged system (e.g., 180 min), were identical within the experimental uncertainty, can be best explained as the result of clustering of the initially nano-sized, individual reverse micelles into large objects which are themselves embedded in a "sea" of non-clustered micelles. This indicates that in the mixed system the total population of individual small micelles remained fixed over time, yet the arrangement is space of the individual micelles changed over time leading to the formation of the large aggregated clusters. The shape of the scattering patterns at $q < 0.2$ nm$^{-1}$ also indicates that after ~ 20 min (arrow in Fig. 1) the size of the individual, large aggregated objects grew very rapidly to diameters >150 nm and thus fell outside the detectable $q$-range *i.e.,* once the intensity started increasing at low-$q$ practically no Guinier plateau in the $I(q) \propto q^0$ region was observed. Hence, the continuous increase in scattering intensity between 0 and 90 min originated primarily from the growing number density of these large aggregated objects and not from their gradual growth in size. This interpretation is further evidenced by the SAXS measurements of CaCO$_3$ microemulsions centrifuged to the point that 98% of the supernatant was removed (see Fig. S2A and Materials and Methods). This procedure was intended to considerably increase the contribution of the large aggregated objects to scattering with respect to the "sea" of the non-clustered micelles. This effect was visible by a shift of the low-$q$ intensity increase region (see the arrow in Fig. S2A) from $q < 0.2$ nm$^{-1}$ (the non-centrifuged case) to $q < 0.4$ nm$^{-1}$ (the centrifuged case). We point out that, at the same time the centrifugation procedure did not affect the form factor part of the profile ($q > 0.4$ nm$^{-1}$), which was identical to the final stage of aggregation of micelles.

*Direct in situ imaging of the gradual formation of the micellar clusters*

To directly image in real space the formation and growth of the large objects (> 150 nm) characterized above from the SAXS measurements in reciprocal space, we employed time-resolved and *in situ* LC-STEM approach. Indeed, in Fig. 2A selected time-resolved images extracted from Video S1, document the formation of globular objects (labeled 1-4 in Fig. 2A), which match the large aggregated objects observed in our SAXS. In the LC-STEM images, the observed globules far exceed in size both the dimensions of the initial individual reverse micelles (diameter ~ 7 nm), and the maximum size observable in SAXS (>150 nm), since in the LC-STEM observations the imaged globules reached a radius of approx. 900 nm several minutes after the onset of their growth (Fig. 2B). Interestingly, in the LC-STEM experiments the formation of the globular structures appears to be significantly faster than in our scattering experiments (Fig. 1). However, with SAXS we measured the overall progress of the reactions, and showed that the change in scattering intensity reflected a combination of the growth rate of individual globules and the global "rate of appearance" of new globules (i.e., the increase of their number density, see also Note S2). The time-resolved LC-STEM images within which we follow 4 tracked objects revealed that these started to form only after mixing of the two microemulsions in the liquid-cell (t = 0 seconds in Fig. 2A). This shows that the globular objects formed through the interaction between the two individual salt-carrying micelles upon mixing. Importantly, the globules were soft, nearly spherical and liquid-like in nature. This is clearly illustrated by the growth of objects 1′ - 1‴ (Fig. 2A), which merged into a new spherical object 1 within a single frame during the *in situ* imaging. The liquid-like properties were further evidenced by the fact that large globules could expand near-instantaneously (within a single frame – Fig. 2B) at the expense of smaller ones. For example, tracked object 4 in Fig. 2A revealed that it expanded its radius after 100 seconds of growth (relative time) by 20% (sharp increase in diameter; Fig. 2B) through the rapid merging with / off several smaller neighboring globular objects. The growth profiles of the individual globular objects yielded comparable sizes as a function of relative growth time (Fig. 2B), indicating the relative homogeneity of the physicochemical conditions within the field of view (Fig. 2A). Considering that the diameters of the globules were in fact larger than the 500 nm spacer between the $Si_3N_4$ windows of the liquid-cell system (Fig. 2 and Methods), leads to the question as to what extent the confined environment between the $Si_3N_4$ windows affected the growth of the globules (see also Note S2)? If the confinement played a major role, the globules should become either lens-shaped, or, depending on the wetting between the globules and the $Si_3N_4$ membrane, be truncated as they grow and reach the opposite-faced membrane. To distinguish between these two possibilities, we correlated the cross-sectional diameters of globules 1-3 (Fig. 2A), with the corresponding transmission intensities corrected for the temporal background levels (Fig. 2C), revealing a linear dependence between globule sizes and intensity. This correlation suggests that the globules remained spherical in shape even if their diameters grew larger than the 500 nm of the spacer and despite the apparent confinement further indicating that the overall fluid layer thickness inside the liquid cell was significantly larger than 500 nm. This can be explained by the fact the $Si_3N_4$ membranes are not stiff and can bow, allowing for membrane-liquid interactions at locally larger distances(28). It is important however, that all globules grew at equivalent rates (Fig. 2B) and reached a similar size plateau with diameters just under ~ 2 μm. We conclude that this value corresponds to the actual local distance between the membranes within the field of view. Thus, it seems that the growth of the globules was clearly restricted upon reaching the hydrophilic surface of the opposite $Si_3N_4$ window (see Methods). Combined, the spherical shape of the globules and their restricted final sizes imply that the contact angle between the globules and the membrane is very high, pointing to a low affinity between the hydrophilic membranes and the developing structures. In turn this confirms that the globular objects are hydrophobic in nature, which is due to the presence of terminating hydrophobic tails from the surfactant at the surfaces of globules (see Materials and Methods and Fig. S1). This matches the expected behaviors for clusters of aggregated reverse micelles as derived above from the SAXS data.

Combining the LC-STEM and SAXS analyses suggests that the globular objects grew by aggregation/clustering of individual reverse micelles. Furthermore, once the globular aggregates of micelles reached a plateau, they stopped growing and remained stable (Figs. 1A & 2B). However and most significantly, in none of the analyses above did we observe any solid or particulate morphologies (e.g. (21, 22, 29–32)) that would suggest the formation of either crystalline $CaCO_3$ or amorphous calcium carbonate (ACC) solid phases. Therefore, based on our observation we can assert that in our mixed microemulsion system we only have no solids, despite the fact that we are highly supersaturated with respect to all $CaCO_3$ phases (see Materials and Methods).

*In situ and time-resolved crystallization*

To assess if and how the as-formed aggregated globular structures can crystallize, we induced a change in the composition of the microemulsion environment through the injection of ethanol into the liquid cell. This lead to an immediate destabilization and disintegration of the globules and the subsequent, almost instantaneous, formation of a solid phase (Fig. 3A, Video S2). We quantified the formation kinetics of this new phase using the cross-sectional area of particles within the field of view as a proxy (Fig. S3A). In the first stage of reaction (up to 50 s), the globules instantaneously disintegrated once they came into contact with the passing alcohol front (compare Fig. 3A at 0 seconds and the decrease in area and change in shape of the globules at 30 and 50 seconds, Fig. S3A). Once disintegrated, the growth of a new phase occurred and it followed a sigmoidal dependence, until nearly the entire field of view was filled with this phase (Fig. S3A). The newly crystallized phase grew into dendritic-like structures. We used *ex situ* TEM analysis of the phase that grew on the $Si_3N_4$ windows and show that two types of features formed (i) large dried and burst globular objects, made of fan/flower like shapes (Fig. S3B), and (ii) large dendritic structures that extended for tens of microns across the window (Fig. S3C). These fan and dendrite morphologies formed under flow conditions and in the confined space of the liquid cell (Fig. 3 and Fig. S3A) and based on the *ex situ* TEM analyses (Fig. S3C) were determined to be crystalline. The nature of this *in situ* formed mineral phase was revealed by selected area electron diffraction to be the $CaCO_3$ polymorph vaterite (Fig. S3D, see also Materials and Methods, Table S2, Note S3, Figs. S4 & S5). The *in situ* SAXS from the $CaCO_3$ microemulsion measured during the analogous ethanol injection in bulk (see Materials and Mehods and Figs.S2B&C) showed that the destabilization at the nanometre-scale involved a total disruption of the micellar structure, where the intensity at $q > 0.2$ nm$^{-1}$ decreased by a factor of ~10 times, as the form factor of the micelles had completely disappeared.

*Do the globular aggregates contain solid calcium carbonate before the induced crystallization?*

Over the course of our investigations the fundamental question arose, whether the mixed microemulsions that formed the globular aggregates contained solid calcium carbonate phases before the globules were destabilized. In order to answer this we performed *in situ* HEXD analyses on the aged (~4 h) $CaCO_3$ microemulsion containing globules (see Materials and Methods). From these HEXD measurements, we derived the pair distribution function $g(r)$ (PDF; Fig. 4A) of the globules, which we compared to the PDFs of various solid $CaCO_3$ phases and the independent components of the microemulsion. The $g(r)$ of the aggregated globular $CaCO_3$ microemulsion (A) was different from the actual microemulsion components (Fig. 4B&C) and showed a characteristic and unique peak at 2.46 Å, which was attributed to a Ca – O distance(33). This value lies between the Ca – O bond distance in the $g(r)$s in solid amorphous calcium carbonate (ACC; 2.39 Å; Fig. 4D) and aragonite (2.51 Å; Fig. 4G). However, and importantly, compared to any of the solid samples analyzed (patterns D to G) the structural coherence in the microemulsion sample (A) did not extend above ~ 4 Å, which is characteristic for liquids (for instance Fig. 4C). Such short coherence suggests that the analyzed mixed and aggregated $CaCO_3$ microemulsion globules contained only solute Ca-$CO_3$ species and not an amorphous (D) or crystalline solid $CaCO_3$ phase (E, F G in Fig. 4). Furthermore, our SAXS characterization of both non-centrifuged and centrifuged aged $CaCO_3$ microemuslions (compare Figs. 1 & S2) points out to no morphological changes to the microemulsion at high-$q$. That means that no species were present, which would have a form or electron density different than the micelles themselves.

**Discussion**

When the reverse micelles collide, due to the attractive forces and the fluidity of the dynamic surfactant films, they fuse to transitional clusters, exchange their water cores' contents (e.g., ions) and then again separate (fusion-fission mechanism(16, 18)). This communication is an intrinsic property of microemulsions. In our system, the inter-droplet $Ca^{2+}$ and $CO_3^{2-}$ exchange enables interaction and exchange between the individual reverse micelles. Such chemical communication and exchange should invariably lead to supersaturation with respect to calcium carbonates inside the water nanoreactors and thus this should trigger precipitation of a solid phase. Why does no solid $CaCO_3$ phase form/precipitate, and why do these micellar communications instead lead to the formation of large and stable aggregated globules?

In our experiments, the diameters of the water cores in the initial, separated reverse micelles were in the range of 3 – 4 nm (Fig. S1). This means that at the initial, nominally high $Ca^{2+}$ and $CO_3^{2-}$ concentration of 0.15 mol/L, each water core contained on average only ~1 – 3 ions of calcium or carbonate (in total ~4.5 – 9 of all added ions, see Note S4). This clearly illustrates, that a single encounter between a $Ca^{2+}$-carrying and a $CO_3^{2-}$-carrying reverse micelles would by no means yield a solid phase, although the concentration of ions was significantly above saturation for ACC, vaterite, calcite or aragonite (see Materials and Methods). On the other hand, it has been recently demonstrated that the calcium and carbonate ions show high mutual affinity and will form stable solute ionic associates in solution upon encounter even at highly undersaturated conditions.(34–37) These associates might be akin to the concept of the pre-nucleation clusters observed by Gebauer et al.(34), but could very well be ion pairs. (37) Regardless of their actual form, either poly-ionic or simpler, it is important to note that up to ~75% of calcium becomes bound in solution before nucleation of the solid phases. (34, 37) Thus, our observations could be explained in terms of a selective accumulation of such small $CaCO_3$ solute species (whatever they are, compare (34) vs. (37)) in the water cores of the reverse micelles. It has been established that in NaAOT-based microemulsion systems even small amounts of ionic species in the water pools can affect the stability and reactivity of reverse micelles(38, 39). Furthermore, with increasing salt concentrations inter-micellar attraction is attenuated, leading to micelles behaving more like hard-spheres, and increasing the stiffness of the surfactant interface(27). Therefore, an increase in the electrolyte concentration should lead to a decrease in the reactivity of the microemulsion environment, and a decreased tendency to form clusters/aggregates of micelles.(40) On the other hand, lower ionic strengths inside the water cores should favor mixing of the water core contents, and formation of large clusters of micelles. In our case, we start with two initial microemulsions, which are stable. The presence of the dissolved ions impedes to a certain degree the overall communication between the aqueous cores upon mixing, yet, the actual ionic exchange processes occurs, but at lower probabilities of effective fusion-fission (Fig. 5A). Since we know that the formation of globules is caused by the inter-micellar communication, the rate of ion exchange should be correlated with the rate of globule growth. In other words, the globules should form faster at lower initial ion concentrations at a constant water-to-surfactant ratio $w$ (i.e., equivalent size and number density of micelles). We tested this hypothesis by following the formation of globules using changes in the optical turbidity of mixed samples as a function of time at three different total ion concentrations (between 0.05 mol/L and 0.1 mol/L with respect to $CaCO_3$) but all at $w$=10. The comparison (Fig. S6) shows that the globules form indeed at faster rates for lower ion concentrations, supporting our hypothesis. Our data suggest that as the ionic exchange progresses through fusion-fission (Fig. 5B), Ca-CO$_3$ solute species form. Consequently, the effective ionic strength in the aqueous cores will decrease as increasingly more ions become associated with each other. Thus, the surfactant interface of micelles containing Ca-CO$_3$ species becomes less stiff, and such micelles likely show higher tendencies to cluster/aggregate into globules (Fig. 5C). This decrease in stiffness and increase in the tendency to form globules is also reflected by a small decrease in the volume of the individual micelles observed during the reaction (5% after 180 minutes; Fig. 1), which is expected for the transition to a more fluid surfactant interface(17). According to the model of solubilization of small molecules(17), the decrease in micellar radius and the decrease of stiffness of the micelle interfaces should occur when the solute molecules are directly anchored to such interfaces. In our system, this results in a population of micelles that do not (practically) contain Ca-CO$_3$, in their water cores (Fig, 5B) but instead primarily contain dissociated Na$^+$ and Cl$^-$ ions (Fig, 5C at right) that contribute to the relatively increased salinity in the water pools. Hence, this second population of micelles is effectively less "sticky" and does not easily form aggregated globules. In experiments that utilize purely inorganic aqueous solutions, amorphous calcium carbonate (ACC) is the first solid phase that precipitates from bulk solution(31, 41) Its formation, stability and subsequent transformation to crystalline CaCO$_3$ phases is highly dependent on variations in physicochemical properties of the systems observed(29–32). In contrast, in our microemulsions we do not have a bulk homogenous solution. Instead each individual microemulsion micelle has a highly confined volume and is less than 4 nm in diameter. Models suggest that highly confined volumes (*e.g.,* nanopores) can contain solutions at higher supersaturations when compared to unconfined bulk conditions (*i.e.,* pore-size controlled solubility effect, where an effective solubility is defined for confined volumes, following the Young-Laplace equation(42, 43)). Furthermore, recent studies(44, 45) have shown that highly hydrophilic substrates are not good templates for heterogeneous nucleation, due to the high energetic barriers associated to surface dehydration. Assuming therefore that the highly hydrophilic inner surfaces of micelles do not provide a good substrate for heterogeneous nucleation, the effective solubility becomes the most important thermodynamic parameter controlling homogeneous nucleation in the micelle-water system. Applying the Young-Laplace equation(42, 43) to a micellar water pool of 4 nm predicts that the effective solubility of $CaCO_3$ (calcite) under such confinement is ~3.5 times higher than the bulk solubility of macro-crystalline calcite. This suggests that within the confinement of our < 4 nm sized micellar water cores and considering the highly limited temporal and spatial availability of ions, the precipitation of a solid inside such water cores is far less favorable than in the bulk. At the same time however, it was shown that very e.g small $CaCO_3$ clusters/species (1.4 nm) can be stabilized by organic ligands such as 10,12-pentacosadiynoic acid(46). This is likely also the reason why in our experiments, the microemulsion environment similarly hinders the formation of solid $CaCO_3$ phases and favors the stabilization of small poly-ion clusters/species. We could induce crystallization only upon the bursting of the globular aggregates following the addition of ethanol (Fig. 3) because this led to the rapid removal of the confinement and to the fast decrease in the stabilizing properties of the micellar interface. The injection of ethanol (Fig. 3 and Video S2) perturbs near-instantaneously the surfactant film and destabilizes the water droplets and hence also the ions inside the globules (see also Figs. S2B&C and S3). This disruption naturally leads to instant phase separation and disintegration of the globules followed by the fast growth of solid $CaCO_3$ from the supersaturated aggregated globular liquid-like $Ca-CO_3$ phase (Fig. 5D).

**Conclusions**

Our results highlight a strong feedback between the changing state of the interacting ionic species in the water cores and the structure/composition/interfaces of the reverse micelles themselves. The proposed stabilization and 'triggered' destabilization mechanisms are perfect examples of a bioinspired nano-scale mediated mineral nucleation and growth process: the organic interface – inorganic species interactions compete with the purely inorganic drivers changing the thermodynamics of our system and stabilizing a solute $Ca-CO_3$ phase at conditions where a solid would have formed in a bulk solution. Only in the presence of a stable micellar interface, the solute $Ca-CO_3$ species could be preserved within clusters/aggregates of reverse micelles. This is in stark contrast to non-micellar fully inorganic aqueous systems, where solids form instantaneously at the same chemical conditions. Finally, only when a perturbation was introduced by ethanol addition, did the micellar interface become unstable, allowing for the unhindered mixing of the aqueous phases and their contents, and consequently lead to vaterite crystallization.

**Materials and Methods**

Detailed Materials and Methods can be found in SI Appendix. The microemulsions consisted of 0.1 mol/L dioctyl sodium sulfosuccinate (the surfactant, NaAOT, >98%), dissolved in 2,2,4-trimethylpentane (the oil, isooctane) and water. Typically in our experiments the molar ratio between water and the surfactant was $w = 10$. Two separate initial microemulsions were prepared containing either 0.15 mol/L $CaCl_2$, or 0.15 mol/L $Na_2CO_3$. To initiate a reaction we mixed equal volumes of the initial reverse microemulsions which yielded a final microemulsion of $[Ca^{2+}] = [CO_3^{2-}] = 0.075$ mol/L. The products of this reaction were studied *in situ* with SAXS, HEXD and LC-TEM.

**Acknowledgments**

This work was made in part possible by a Marie Curie grant from the European Commission in the framework of the MINSC ITN (Initial Training Research network, project number 290040) and the financial support of the Helmholtz Recruiting Initiative to LGB. We acknowledge funding by the the UK Engineering and Physical Sciences Research Council (EPSRC) (Grant No. EP/I001514/1), funding the Material Interface with Biology (MIB) consortium to RK. We thank Diamond Light Source for access to beamline I22 (experiment number SM8742) and beamline's staff, A.J. Smith and N.J. Terrill, for their competent support and advice during the collection of the small-angle X-ray scattering data. The high-energy X-ray diffraction experiments were performed on beamline ID15B at the European Synchrotron Radiation Facility (ESRF), Grenoble, France (experiment number CH4155) and we thank A. Poulain for assistance during beamtime. This research was partially made possible by Marie Curie grant from the European Commission in the framework of NanoSiAl Individual Fellowship (project number. 703015) to TMS.

**Figure Legends**

**Fig 1.** *In situ* and time-resolved SAXS patterns from a mixed CaCO$_3$ microemulsion. The first 8 frames (up to 2 minutes and 40 seconds) correspond to a pure CO$_3^{2-}$-microemulsion, after which a Ca$^{2+}$-carrying microemulsion was injected. Between the moment of mixing and up to ~ 20 minutes (arrow) no difference between the single-component system and the mixed system was observed; following this the gradual increase in intensity at $q < 0.2$ nm$^{-1}$ up to 90 min indicates the formation of large scattering objects; furthermore, at $q > 0.2$ nm$^{-1}$ changes in the scattering patterns (intensity and shape; see inset $I(q)q^4$ *vs.* $q$ time resolved plot) translated into a minor (~ 1.5%) decrease in an average micellar radius (see arrows in inset pointing to the shift in position of the first minimum in the plot ($q_{min}$) towards higher-$q$, where the total average radius of the micelle is ~4.5/$q_{min}$). This change translates into a ~5% volume reduction of the micellar water cores when the constant size of the surfactant polar heads is taken into account.

**Fig. 2.** The evolution of globules as observed in LC-STEM. The images were obtained at a low electron beam current (4 pA), which minimized the electron dose reducing it by orders of magnitude compared to plane-wave illumination in TEM. This allowed us to image at low dose values of 3 e/nm$^2$ at the chosen magnification, and enabled us to minimize electron beam related artefacts. In the field of view, an individual pixel corresponded to approx. 14 x 14 nm$^2$, thus not allowing for the imaging of individual reverse micelles. We observed the formation of large globules only following the mixing of the Ca$^{2+}$- and CO$_3^{2-}$ carrying microemulsions inside the liquid-cell, and never when we only imaged the initial non-mixed salt-carrying microemulsion. A-C are derived from Video S1; in A the time scale corresponds directly to that of the video's, whereas in B the time scale represents a relative growth periods as each of the tracked objects, which started developing after certain induction periods. A) Selected unprocessed images from the growth sequence of several globules marked 1-4; object 1 was formed from three smaller units 1′-1‴ that grew independently of each other until the 50$^{th}$ second; objects 2, 3 and 4 grew gradually (except that object 4 increased in size by 20% within 1 frame after 100 s of its relative growth time – see B); interestingly all tracked globular objects reached similar maximum sizes of ~ 1600 nm B) change in diameters of tracked objects 1-4 as a function of time; C) correlation between diameters of objects 1-3 from the selected time series, and the recorded transmission intensity; colored symbols and lines used in B&C, correspond to the objects 1-4 indicated by the same colors in A.

**Fig. 3.** Destabilization of the CaCO$_3$ microemulsion. Selected unprocessed images from the destabilization sequence of globules as observed by LC-STEM upon *in situ* ethanol injection into the cell taken at 3 e/nm$^2$; the purple arrow in the frame at 50 s points to the contrast change caused by the passing ethanol front; note however, that the burst of the 1$^{st}$ globule was already visible at < 30 s; the dynamics of the reaction is better seen in Video S2; The time scale correspond to that in the Video S2, where t = 0 s marks the injection of ethanol after the image sequence in Video S1 had been taken.

**Fig. 4.** Atomic PDFs of the aggregated globular CaCO$_3$ microemulsion compared to microemulsion components and standard solid CaCO$_3$ polymorphs. A) a centrifuged 0.075 mol/L CaCO$_3$ microemulsion at *w*=10 (with isooctane background subtracted); the *g*(*r*) shows a characteristic unique peak at 2.46 Å and no structural coherence above ~ 4 Å. The peak can be attributed to a Ca – O distance(33); B) surfactant component of the microemulsion: 0.1 mol/L NaAOT solution in isooctane with the medium subtracted as a background; C) oil component of the microemulsion - pure isooctane; D) solid amorphous calcium carbonate (ACC); E) crystalline vaterite; F) crystalline calcite; G) crystalline aragonite; PDFs D-G were measured using dry powder samples. A characteristic "rippling" in patterns A to C at *r* > 4 Å is an artifact of the Fourier transform of the as-collected HEXD data and is a consequence if decreased signal-to-noise ratios in liquid samples compared to dry powdered samples shown in D-G.

**Fig. 5.** Schematics of the microemulsion processes leading to the formation of globules followed by the induced solid CaCO$_3$ formation. A) Initial state and mixing of two ion-carrying microemulsions. In NaAOT-based microemulsion systems(17, 40) where salt ions are present the chemical exchange/communication between the aqueous ion-containing water-cores is partially impeded. In general, at raising salt concentrations, the effective polar head area of the NaAOT surfactant decreases, due to the screening by any additional ions, which also leads to a decrease in the repulsion between the heads. Therefore, the presence of salts causes the surfactant packing parameter(40) *v/al* (where *v* is an effective volume of a NaAOT molecule, *l* length of the nonpolar tail) to increase. This salt-induced increase in *v/al* translates into a simultaneous increase in the stiffness of the surfactant interface and this in turn hinders percolation and ion exchange between the individual micelles. As a consequence, any opening of inter-droplet channels and exchange of water core contents is less likely(40). On the other hand, lower salinities (relative to starting conditions) favor "sticky" dynamic interactions, and the formation of large micellar globular aggregates; B) inter-micellar interaction and ion-exchange, driving the formation of stable liquid-like Ca-CO$_3$ species (possibly akin to ion pairs or prenucleation clusters(34)); C) formation of clusters of micelles containing stabilized Ca-CO$_3$ species (i.e., low salinity and hence "sticky"), surrounded by a "sea" of hard-sphere-like micelles containing primarily dissociated counter ions (i.e., high salinity); D) destabilization through the break-down of the surfactant interface due to the presence of ethanol, followed by the unrestricted merging of water core contents, and the rapid association of small Ca-CO$_3$ species in the aqueous phase leading to the precipitation of solid CaCO$_3$. The aggregates in C and D are schematic, since the actual observed morphologies (Fig. 2) include 10s of thousands of aggregated/clustered micelles.

*Supporting Information Appendix*

*for*

*"On demand" triggered crystallization of $CaCO_3$ from solute precursor species*

Tomasz M. Stawski[1,2], Teresa Roncal-Herrero[3,4], Alejandro Fernandez-Martinez[5], Adriana Matamoros Veloza[2], Roland Kröger[3], Liane G. Benning[1,2,6]

[1]German Research Centre for Geosciences, GFZ, Interface Geochemistry, Potsdam, Germany; [2]School of Earth and Environment, University of Leeds, Leeds, UK; [3]Department of Physics, University of York, York, UK; [4] Institute of Microwaves and Photonics, University of Leeds, Leeds, UK; [5]Université Grenoble Alpes, CNRS, ISTerre, Grenoble, France; [6]Department of Earth Sciences, Freie Universität Berlin, Berlin, Germany.

**Corresponding authors**:

Tomasz M. Stawski; German Research Centre for Geosciences, GFZ, Interface Geochemistry, 14473, Potsdam, Germany; +4933128827511; stawski@gfz-potsdam.de

Roland Kröger; Department of Physics, University of York, YO10 5DD, York, UK; +441904324622; roland.kroger@york.ac.uk

Liane G. Benning; German Research Centre for Geosciences, GFZ, Interface Geochemistry, 14473, Potsdam, Germany; +4933128828970; benning@gfz-potsdam.de

**Materials and Methods**

*Microemulsion preparation*

All experiments were carried out with water-in-oil reverse microemulsion that were equilibrated at 20 °C prior to use. The microemulsions consisted of 0.1 mol/L dioctyl sodium sulfosuccinate (the surfactant, NaAOT, >98% Sigma Aldrich), dissolved in 2,2,4-trimethylpentane (the oil, isooctane, HPLC grade, Fisher) and water. Typically in our experiments the molar ratio between water and the surfactant was $w$ = [$H_2O$]:[NaAOT] = 10. Two separate initial microemulsions were prepared with water (cores containing either 0.15 mol/L $CaCl_2$ (hexahydrate, >99%, Fisher), or 0.15 mol/L $Na_2CO_3$ (anhydrous, >99.9%, BDH Chemicals Ltd.; ultrapure deionized water, MilliQ, resistivity >18 MΩ cm). All the chemicals were used as purchased and without any further purification. The individual, initial unmixed microemulsions, were optically clear and stable for extended periods of time (weeks), but for all experiments fresh microemulsions (maximum 3-4 hours old) were prepared. The chemical compositions of the reverse microemulsions used corresponded to a region of the phase diagram(47), where only reverse micelles are stable. We also confirmed the stability of the individual unmixed microemulsions by small-angle X-ray scattering analyses (SAXS, see below). Using a $w$ = 10 and the above mentioned salt concentrations resulted in individual microemulsions considerably below their water (brine) solubilization capacity ($w_{max}$ ~ 20, at [salt] = 0.15 mol/L, see also Note S1). To initiate a reaction we mixed equal volumes of the initial reverse microemulsions containing $Ca^{2+}$ and $CO_3^{2-}$ ions respectively. The mixture yielded a microemulsion of [$Ca^{2+}$] = [$CO_3^{2-}$] = 0.075 mol/L. Over the course of ~3 hours and under gentle stirring at 150 rpm, a translucent whitish colloidal suspension developed and we followed the development of this suspension with time. Theoretically, upon inter-micellar ion-exchange, at the local $[Ca^{2+}] = 0.075$ mol/L and $[CO_3^{2-}] = 0.075$ mol/L concentrations, the aqueous phase should become supersaturated with respect to calcite (supersaturation index, $SI_{Cc}$ = 4.00), aragonite ($SI_{Ar}$ = 3.85), vaterite ($SI_{Vat}$ = 3.37) and amorphous calcium carbonate, ACC ($SI_{ACC}$ = 1.78), as calculated with the geochemical computer code PHREEQC(48). Please note that within the bulk composition of the mixed microemulsions, the water cores constituted only 1.77% of the total microemulsion volume fraction, and thus the actual corresponding total maximum possible yield of $CaCO_3$ was ~1.3 mmol/L. However, this value is obviously irrelevant in terms of the calculation of supersaturation indices above since, they have to be calculated assuming only the aqueous phase volume and not the total complex system, since $CaCO_3$ is insoluble in isooctane.

Prior to choosing the $w$=10 and salt concentration of 0.15 mol/L as the most adequate system, we tested several other compositions with respect to $w$ at a constant surfactant concentration (before mixing): $5 \leq w \leq 15$, and $0.05 \leq$ [salt] $\leq 0.20$ mol/L. Apart from $w$ = 15 and $[CaCl_2]$ = 0.20 mol/L in all cases stable, single-phase optically-clear microemulsions formed. Upon mixing in all experiments again stable microemulsions formed and no variability in the reaction products was observed. We decided to use the $w$=10 and 0.15 mol/L ion concentration system as this yielded the highest possible amount of $CaCO_3$, while remaining safely within the microemulsion stability region (solubilization capacity). Furthermore, we also tested several other compositions outside the stability region (see also Note S1) to optimize the system as best as possible. These tests also included analyses using *in situ* and time-resolved turbidity measurements by UV-Vis spectroscopy (Uvikon XL) to determine the spectral features, which could be associated with the processes upon mixing of the two initial microemulsions (Fig. S6A). These tests revealed that the progress of reaction could be monitored through the evolution of the local maximum at ~275 nm (measured after 6 h). We have hence measured changes in absorbance as a function of time for three microemulsions of $w$=10 and different salt concentrations (Fig. S6B) and used these as a proxy for reaction progress. All the UV-Vis measurements were performed in a quartz cuvette with UV-VIS data collected every second.

*Small-angle X-ray scattering (SAXS) characterization*

All *in situ* and time-resolved SAXS measurements were carried out at beamline I22 of the Diamond Light Source Ltd (UK). The microemulsions were continuously stirred at 150 rpm in a 200 mL glass reactor, and circulated through a custom-built PEEK flow-through cell with embedded borosilicate capillary aligned with the X-ray beam (ID 1.0 mm, wall thickness ~10 μm) using a peristaltic pump (Gilson MiniPuls 3, flow ~10 mL/minute). To initiate the reactions we used a secondary peristaltic pump (Gilson MiniPuls 3) as a remote fast injection system. Typically, 15 mL of a $Na_2CO_3$-containing microemulsion were circulated through the cell and 8 scattering patterns were collected at a rate of 20 s/frame to obtain data for the initial, unmixed microemulsions. While continuing the scattering data collection, we injected 15 mL of a $CaCl_2$-containing microemulsion remotely into the reactor. The injection lasted for ~ 20 s, and no changes in the measured scattering pattern intensities or shapes were observed during injection. The development of the microemulsion mixtures were followed up to ~3 hours with patterns acquired every 20 seconds. Additionally, we also performed an experiment at a time-resolution of 1 second, in which we injected 10 mL of ethanol (96%) into 50 mL of aged (~4 h) $CaCO_3$ microemulsion to study the processes of its destabilization.

The SAXS measurements were performed using a monochromatic X-ray beam at 12.4 keV and two-dimensional scattered intensities were collected at small-angles with a Dectris Pilatus 2M (2D large area pixel-array detector). Transmission was measured by means of a photodiode installed in the beam-stop of the SAXS detector. A typical sample-to-detector distance of 4.27 m allowed for a usable $q$-range of $0.038 < q < 3.70$ nm$^{-1}$. We also performed several measurements at 1.2 m and a $q$-range $0.100 < q < 7.00$ nm$^{-1}$ in order to study fine changes in the structure of micelles. The scattering-range at small-angles was calibrated against silver behenate and dry collagen standards. The scattered intensity was calibrated to absolute units against the 1 mm glassy carbon standard available at I22. The reactions described in the "Microemulsion preparation" section were followed *in situ* from the very early stages and up to the point when no changes in sample transmission were observed (a direct readout from the photodiode) at the time resolution of 20 s/frame. For each of the experiments we measured a series of backgrounds and reference samples, which included: the empty cell, isooctane background, pure-water microemulsion at $w = 10$. All recorded 2D SAXS patterns were isotropic and could therefore be reduced to 1D curves. Background subtractions, data corrections, normalizations, and integrations to 1D curves were performed using the DAWN software (v. 1.3 & 1.4) according to I22 guidelines.

*In situ liquid-cell STEM/TEM characterization (LC-STEM)*

To image the development of the microemulsions containing Ca$^{2+}$ and CO$_3^{2-}$ ions upon mixing, a Protochips Poseidon 200 liquid cell holder was employed. This holder has two injection inlets and one outlet port. The imaging was carried out through two 50 nm thick silicon nitride membranes with a total accessible area of 550 µm x 50 µm separated by a 500 nm spacer. However, it is important to note that the effective fluid thickness is significantly larger due to membrane bowing effects(28). The membranes were chemically cleaned prior to assembly by rinsing with acetone and ethanol (Sigma Aldrich, ACS reagent grade). To make the silicon nitride membranes hydrophilic a 50 W oxygen plasma was applied for 2 min (Femto, Diener Electronic GmbH). 40 x 10 nm gold nanorods (Strem Chemical Inc.) were deposited as fiducial markers onto the top membrane prior to a second plasma treatment. The chips were assembled using a 0.5 µL droplet of the 0.1 mol/L NaAOT solution in isooctane sandwiched between them to prevent membrane collapse. A pure, 0.1 mol/L NaAOT solution in isooctane was connected to one liquid cell inlet via a 3 mL syringe and PEEK tubing with an internal diameter of 100 µm. The NaAOT solution was continuously pumped through the cell by a syringe pump (Harvard Apparatus Inc.) at a flow rate of 300 µL/h. Once a stable and focused image was obtained, using the gold nanorods as fiducial markers, a $Ca^{2+}$-containing microemulsion was pumped through one of the inlets and the $CO_3^{2-}$-containing microemulsion was injected through the second inlet. Both emulsions mixed in the tip volume of the holder. Additionally we also imaged *in situ* each of the salt-carrying microemulsions to check for any presence of globules, and we confirmed that no such morphologies were present, and hence they formed only upon mixing of the two microemulsions. In order to destabilize the as-formed globules and to induce the crystallization of $CaCO_3$, the flow of the ion-carrying microemulsions was stopped, and in their place ethanol (96%) was pumped into the cell at a flow rate of 300 µL/h. For the analysis and *ex situ* imaging of the mineral phase, which formed after the ethanol injection, the liquid-cell holder was disassembled and the membrane was analyzed using conventional TEM (see below).

To image the mixing and microemulsion reaction dynamics in the LC-STEM cell, a double aberration corrected JEOL JEM 2200FS (S)TEM operated at 200 kV using a high-angle annular dark-field scanning TEM (HAADF-STEM) detector was used. A 12 mrad α semi-angle condenser aperture was used and the lateral size of the probe was approx. 1Å. Images were recorded employing a K2 camera of a physical resolution of 1024 x 1024 pixels$^2$ at a dwell-time of 0.5 μs per pixel resulting in a scan time per frame of approx. 0.5 s. Images were recorded at 20 k magnification resulting in a single pixel area of 14 x 14 nm$^2$. Videos were acquired using the software Camtasia allowing for the capture of live-view images using Digital Micrograph (JEM 2200, JEM 2011) and were down-sampled to 512 x 512 pixels$^2$ and stored at 2 fps (see Videos S1 and S2 with real-time data; Videos S3 and S4 contain the same data but are sped up 10 times to 20 fps). The time-scales in Figs. 2&3 in the main text are independent from each other, and in each case 0 seconds marks the onset of a different event. In Fig. 2A "0 s" at the time scale relates to the mixing of the initial microemulsions, in Fig. 2B it marks the relative onset of growth of the tracked objects, whereas in Fig. 3A-B it relates to the initiation of the injection of ethanol into the port of the liquid cell holder. For the post reaction *ex situ* TEM analysis of the mineral phases, a JEOL 2011 was used at a beam current of approx. 5 μA and an acceleration voltage of 200 kV.

For the *ex situ* characterization of the reaction products presented in Fig. S5 and Note S3, the mixed $Ca^{2+}$- and $CO_3^{2-}$ –carrying microemulsions were directly cast onto holey carbon Cu grids (CF200-Cu, Electron Microscopy Sciences). The as deposited and dried samples were analyzed in TEM mode (high-resolution imaging and selected area electron diffraction, SAED) using a FEI Tecnai TF20, with images recorded by a GatanOrius SC600A CCD camera. We characterized the products of reaction of several compositions of the microemulsions with varying *w* and using variable initial ion-concentrations, and we did not observe any differences in morphologies of the particles formed. Therefore, all the presented images are from the 0.075 mol/L $CaCO_3$ system of $w = 10$ after ~3 hours from mixing.

*High-energy X-ray diffraction (HEXD) characterization*

The HEXD measurements were performed on beamline ID15B at the European Synchrotron Radiation Facility (ESRF, Grenoble, France). In order to achieve better signal to noise ratios for the HEXD measurements we centrifuged the mixed but not destabilized microemulsions at 16000 g. Various times were used to evaluate the influence of centrifugation onto the stability of the microemulsion suspension (5 – 120 min), and typically a time of 15 minutes was used. We did this in order to enrich the microemulsion with respect to the Ca-$CO_3$-containing globules by increasing their volume fractions. During centrifugation, the system visibly separated, and 98% of the supernatant could be decanted. The structures of this separated supernatant and of the resulting centrifuged solution as well as the centrifuged initial unmixed microemulsions were also tested with SAXS (at I22, see the SAXS measurement description) to evaluate if centrifugation changes the structure. This analysis confirmed that the post-centrifugation supernatant from the mixed microemulsions contained only reverse micelles but no large scattering objects, and the high-*q* part of the data ($q > 0.2$ nm$^{-1}$) for the centrifuged samples indicated that the structure of the reverse micelles was not affected by the centrifugation procedure (see Fig. S2). The as-prepared samples were transferred into 1 mm Kapton capillaries, and HEXD analyses were carried out at the wavelength of 0.14383 Å allowing us to collect scattering in the *q* range up to ~250 nm$^{-1}$. The 2D HEXD patterns were collected using a MARCCD165 detector (pixel size 150 x 150 μm$^2$) in Debye-Scherrer geometry according to the ID15B guidelines. The sample-to-detector distance was 234.4 mm and was calibrated using a CeO$_2$ standard in 1 mm Kapton capillary. Due to the expected relatively low signal-to-background from the measured samples, we used long counting times (up to 200 s, typically 100 s) and collected multiple frames (up to 300 per sample, typically 20), which were later averaged together. The detector dark-current image was measured every 5 frames and dark-current contributions were corrected automatically by the acquisition software. Furthermore, various background and reference samples were measured under similar conditions, and included among others, an empty Kapton capillary, a capillary containing pure isooctane, and powders of solid CaCO$_3$ phases. The as obtained 2D images were transformed into 1D curves using Fit2D from which the atomic pair distribution functions (PDFs) were obtained by using PDFgetX3 software package(49). The background and Compton scattering subtractions were also performed using the same software package.

**Supporting Video Still Image captions**

**Video S1.** Real-time video showing the evolution of globules as observed in LC-STEM– see also Fig. 2 in the main text. We observed the formation of large globules only following the mixing of the $Ca^{2+}$- and $CO_3^{2-}$ carrying microemulsions inside the liquid-cell, and never when we only imaged the initial non-mixed salt-carrying microemulsion. Video was acquired using the software Camtasia allowing for the capture of live-view images using Digital Micrograph and were down-sampled to 512 x 512 pixels$^2$ and stored at 2 fps.

**Video S2.** Real-time video showing the destabilization of the $CaCO_3$ microemulsion. as observed by LC-STEM upon *in situ* ethanol injection into the cell – see also Fig. 3 in the main text. Video was acquired using the software Camtasia allowing for the capture of live-view images using Digital Micrograph and were down-sampled to 512 x 512 pixels$^2$ and stored at 2 fps.

**Video S3.** The same as Video S1 but sped up to 20 fps.

**Video S4.** The same as Video S2 but sped up to 20 fps.

**Supporting Figure captions**

**Fig. S1. Structure of the reverse micelles from SAXS**. A) Schematic of a reverse micelle as a core-shell aggregate with the surfactant molecules surrounding the water pool. The surfactant's polar heads are located at the interface with the water and they form a shell of a constant thickness $T_s$. In our convention, the water core has a shape of an ellipsoid of a principal core radius $a$, and an equatorial core radius $b$; B) experimental scattering profiles of the $CaCl_2$- and $Na_2CO_3$ -carrying microemulsions of $w$=10 and with salt concentrations equalling 0.15 mol/L , together with fitted curves based on the polydisperse core-shell ellipsoid model (see A), in which $T_s$ was constant and given (see Table S1), and $a$ and $b$ were allowed to be affected by the size distribution(s) between 0.5 nm and 100 nm. The resulting form-independent size distributions from the Monte Carlo fitting with MCSAS(50) are shown in: C) for $CaCl_2$-carrying microemulsion and D) for $Na_2CO_3$-carrying microemulsion. From C and D one can see that both microemulsions had practically the same structure with an average mean radius of the water cores of ~1.7 nm.

**Fig. S2. Scattering profiles of the uncentrifuged, centrifuged, and destabilized $CaCO_3$ microemulsions.** A) The centrifugation processes increases the content of aggregated micelles with respect to the non-clustered ones, which is visible by the shift of the low-$q$ intensity increase region towards higher-$q$ (see the arrow). The blue curves were measured at a sample-to-detector distance shorter than the remaining two, hence they cover a slightly different $q$-range, but the low-$q$ increase region in the blue curves actually extends further to lower scattering vector magnitudes. The intensity in the solid blue curve is scaled with respect to the pink one by a factor of 0.776 times in order to emphasize the fact that the high-$q$ parts of the data overlap perfectly between the pink and blue curves which indicates that internally the aggregates are composed of the same kind of micelles (the original scattering pattern: dashed blue line). The destabilization of the $CaCO_3$ microemulsion occurring upon the ethanol addition; B) comparison of the aged stable and destabilized $CaCO_3$ microemulsion, where the original form factor of the micelle at $q > 0.2$ nm$^{-1}$ (the cyan line) diminishes and completely changes the profile upon the ethanol addition (the pink line) ; the dashed lines indicate characteristic scattering exponents; C) the *in situ* and time-resolved representation of the transition highlighting the changes taking place in the form factor of the micelles; the time axis in reversed.

**Fig. S3. Destabilization of the CaCO$_3$ microemulsion.** A) Time-resolved change in the area of the precipitates in the field of view. B to D: *Ex situ* TEM images and analyses of precipitates formed post ethanol injection on the Si$_3$N$_4$ membrane: B) burst globules revealing fan-flower shaped internal contents; C) dendritic fan-shaped morphologies; D) selected area electron diffraction (SAED) from C; Based on the diffraction spots in SAED (pink numbers) and the corresponding *d*-spacing values (white numbers) we identified the precipitated phase as vaterite (Table S2, Fig. S4). Such a diffraction pattern of smeared individual spots (*i.e.,* mosaicity) is best attributed to a structure composed of smaller units, which are preferentially orientated but slightly misaligned with respect to each other. Similar effects can be seen when microemulsion were destabilised *ex situ* during simple drying (see Note S3 and Fig. S5). The time scale in A, correspond to that in the Video S2, where t = 0 s marks the injection of ethanol after the image sequence in Video S1 had been taken.

**Fig. S4. Simulated diffraction pattern for calcite and vaterite.** The patterns were simulated from the structural information summarized in Table S2.

**Fig. S5. Solid phases formed post drying (and imaging) of CaCO$_3$ microemulsions.** A) TEM micrograph of a dry product obtained from a 0.075 mol/L CaCO$_3$ *w*=10 microemulsion harvested after 3 hours after mixing of the individual microemulsions; inset I shows the SAED from the imaged area with a diffraction pattern clearly matching that of vaterite; B) High-resolution micrograph from an aggregate formed under the same conditions as that imaged in A; inset I shows the crystallization induced by the electron beam as illustrated by the sequence of two images from exactly the same area of the aggregate showing more pronounce lattice fringes after longer exposure as documented by the respective fast Fourier transform (FFT) of the two images, with the second image showing higher intensities of the reflections; inset II shows the FFT from the area enclosed by the square (purple dashed line) in B indicating that lattice fringes match those of vaterite.

**Fig. S6. UV-VIS characterization of the microemulsions.** A) The UV-VIS spectrum of a mixed 0.075 mol/L CaCO$_3$ microemulsion of *w* = 10 reacted for 6 hours compared with the selected microemulsions components (as indicated in the legend); these measurements were used to determine the local maximum absorbance at 275 nm. We associated this maximum with the presence of globules (see the main text) and used it as a proxy for the progress of their formation; B) Normalized absorbance as a function of time for three microemulsions of constant *w* = 10, but with variable [CaCO$_3$] concentration as indicated in the legend.

**Supporting Text**

*Note S1. Structure of initial reverse micelles as derived from SAXS measurements.*

All the morphological descriptions in this SI note consider explicitly the composition a water-in-oil microemulsion in a tertiary system of aqueous phase/NaAOT/isooctane as described in Methods. A complete phase diagram of the system can be found elsewhere(47). In water-in-oil microemulsions the minority aqueous phase (polar phase) is stabilised in isooctane (apolar "oil") by an ionic surfactant (NaAOT) in the form of nm-sized droplets (*i.e.* the reverse micelles, Fig. S1A). This stabilization is due to the structure of the surfactant, where a NaAOT molecule has a polar, hydrophilic primarily Na-sulfonate "head" group, which interacts with the polar aqueous phase, and two apolar, hydrophobic organic "tails", which interact with the apolar isooctane.

We fitted the SAXS patterns of the individual initial non-mixed microemulsions with a core-shell ellipsoid model(25) (Fig. S1B and Table S1). In this model the low polydispersity water core (with the axes *a* and *b*) is surrounded by a shell of monodisperse thickness composed of a monolayer of polar "head" groups of the AOT surfactant (thickness $T_s$)(25). The remaining NaAOT "tails" are ignored in the structural model since their electron density closely matches that of the surrounding isooctane medium and hence does not contribute to the scattering contrast (Table S1). The size and the geometry of the water cores in a microemulsion nanoreactor depend on the water-to-surfactant ratio *w*, and can be therefore change/tuned accordingly(17) within the solubilisation capacity of the microemulsion (*i.e.*, maximum *w* at which phase separation in the microemulsion and an excess aqueous phase is observed, $w_{max}$)(47,51,52) . Important to note is the fact, that although the concentration and speciation of salts dissolved in the aqueous phase have a limited impact on the size of the water pools, their presence considerably decreases $w_{max}$. In Fig. S1B, we show the scattering curves from two microemulsions with $w=10$, but containing different salt concentrations dissolved in the water cores. The two different scattering curves are identical at $q > 0.2$ nm$^{-1}$, but show a difference in intensity (~15%) at low-$q$. This small difference is most likely caused by the presence of a minor population of tiny clusters of dimers or trimers in the $Na_2CO_3$-carrying micelles. As we further corroborate through the good fits with the structural model (Figs. S1B-D), in the two used microemulsions, the water droplets have the same size and shape. On the other hand, as we mentioned the solubilisation capacity of the two microemulsions is dependent on the concentration and nature of the counter-ion salts: our tests showed that at $w = 10$, the most adequate experimental conditions at which the microemulsion was stable, were at a maxmimum concentration of $CaCl_2$ of 0.2 mol/L. In contrast, for the $Na_2CO_3$ microemulsions, concentrations even higher than 1.0 mol/L could be used, without any visible phase separation. Conversely, for a constant concentration of salts (as used in the experiments *i.e.,* 0.15 mol/L), for $CaCl_2$ we could achieve a $w_{max} \sim 18$, whereas for $Na_2CO_3$ $w_{max} > 20$.

The reverse micelles exhibited relatively low polydispersity, but were not fully monodisperse. As previously demonstrated(18) for reverse micelles one has to consider simultaneously two types of polydispersity that could affect the size of the water cores: (1) polydispersity in the number of the NaAOT molecules building up a micelle, (2) shape polydispersity, when the ellipsoidal micelles cyclically "pulse" between the oblate and prolate forms, while preserving a constant volume of the water cores. Although, the analytical equations describing the scattering from such a system and including both kinds of polydispersity are available(18), the actual fitting with those equations does not yield satisfactory results. This is because an *a priori* assumption about the type of distribution(s) would have to be made, and one has to deal with mutual dependencies between fitting parameters. Therefore, we employed an alternative approach in which we used a Monte Carlo fitting procedures (Fig. S1B) implemented in the MCSAS software(50). Our goal was to obtain the model-independent size distribution of the dimensions of the formed reverse micelles (Figs. S1C&D). This way we specified and constrain only the geometrical model of the micelle (core-shell ellipsoid) but derived the required structural information (details Table S1). In Figs. S1C&D we present the resulting histograms representing the size distributions of the dimensions of the ellipsoidal micellar water cores for the $CaCl_2$- and $Na_2CO_3$ -carrying microemulsions used in our experiments.

*Note S2. Apparent growth rates of $CaCO_3$ globules*

By analyzing and comparing the apparent growth rates derived from our *in situ* and time resolved LC-STEM and SAXS data we evaluated the influence of the confinement and electron beam exposure on the evolution of the globules shown in Fig. 1 (SAXS) and Fig. 2 (LC-STEM) in the main text. Although such a temporal and spatial juxtaposition is often difficult because of the differences in the probing volumes accessible to each the two techniques (1-100 $nm^3$ for STEM and 10 $\mu m^3$ - 1 $mm^3$ for SAXS), our data sets allowed us to carry out a comparison. Firstly, the scattering features in our SAXS data that correspond to the individual micelles and those that correspond to the growing globules are reflected in two highly separate regimes in $q$ (low-$q$ < 0.2 $nm^{-1}$ and high-$q$ > 0.2 $nm^{-1}$; Fig. 1 main text). The contrast is clear because even a small number of larger objects will strongly affect the scattering at low-$q$, because the intensity scales with the square of the volume of such scatterers. Thus, the shape of the scattering patterns at $q < 0.2$ nm$^{-1}$ after ~ 20 min (which corresponds with the observable onset of globule growth; see arrow Fig 1 main text) indicates that the size of the growing objects fell outside the $q$-range throughout the entire growth process *i.e.,* once the intensity started increasing at low-$q$ typically no Guinier region was observed. This indicates that the individual large scatterers (the globules) were growing in size very rapidly to a diameter >150 nm and beyond. Hence, the continuous increase in scattering intensity throughout the whole observation period (up to 180 min) originated primarily from the growing number density of large scatterers and not their gradual growth in size. This matches with our LC-TEM observations (Fig. 2 in the main text) that revealed that the growth of the globules, once visible in the images, proceeded extremely fast (within 180 seconds once visible). Nevertheless, the comparison between LC-STEM and SAXS data also showed that the overall evolution in the LC-STEM experiments occurred orders of magnitude faster than in the SAXS experiments (< 5 min *vs.* ~1.5 hours). Hence, although the growth rate of the individual large objects/globules was very rapid in both experiments (a few minutes *per* globule), inevitably the difference in temporal evolution must be attributed to the "rate of appearance" of the individual new globules in SAXS *vs.* LC-STEM. We cannot exclude the possibility that in the LC-STEM images, the high "rate of appearance" of new globules could be related to electron beam-sample interactions or confinement. In all our LC-STEM work we kept the electron dose as low as possible and at least visually no obvious interactions between the growing globules and the electron beam or changes in the globule behaviors were observed. On the other hand, the confined sample environment used for the *in situ* liquid cell microscopic imaging was characterized by a high surface-to-volume ratio in contrast to the bulk *in situ* scattering measurements. Both confinement and beam-interactions at the small scale could promote a more effective formation of hydrophobic globules in the LC-STEM set up compared to the bulk scattering experiments. These could thus explain the observed difference between the two characterization methods.

*Note S3. Ex situ imaging and analysis of the morphologies of dried $CaCO_3$ microemulsions.*

*Ex situ* TEM images of the bulk $CaCO_3$ microemulsion that were reacted for 3 hours and that were deposited onto a TEM grid and washed with pure isooctane to remove an excess surfactant (see also Methods) are shown in Fig. S5. The image reveals sub-500 nm in diameter agglomerates composed of smaller primary particles of 10-15 nm in diameter. The selected area electron diffraction (SAED) from these particles indicated that the agglomerates were composed of nanocrystalline vaterite (inset I in Fig. S5A see also Fig. S4). High-resolution images (Fig. S5B), suggest that all individual primary particles within the agglomerates had similar crystallographic orientations, as also observed during the *in situ* crystallization (Fig. 3 in the main text). This can be deduced from the continuous lattice fringes extending across different individual primary particles (inside the box in Fig. 3B). Important to note is the fact that in contrast to the LC-STEM imaging, in these HR-TEM images we see crystallization that was most likely induced due to drying and / or due to the high vacuum conditions of the TEM column and/or to a certain extent due to the interaction with the beam during the actual imaging. The latter effect was evidenced by the increase in contrast of the lattice fringes, when we compared consequent images from the same area and their FFTs (inset I in Fig. S5B). The *d* spacing of the lattice fringes corresponded to vaterite (inset II in Fig. S5B, Fig. S4, Table S2).

*Note S4. Number of ions per water core*

In order to prepare a microemulsion of $w$=10 from 0.1 mol/L NaAOT in isooctane, we typically mixed 0.18 mL of the aqueous phase/10 mL of the surfactant solution in oil. The mean radius of the water core was ~1.7 nm (see Note 1), hence for the range of radii from 1.5 - 2 nm the volume of the individual water cores was ~14.1 – 33.5 nm$^3$. By knowing the total volume of the aqueous phase and the volume of an individual micelle, we could calculate the total number of micelles in our system (0.18 mL of water) to be ~ $0.54 \cdot 10^{19}$ - $1.27 \cdot 10^{19}$. In this system, the total number of moles of each salt in a 0.15 mol/L solution was accordingly equal to $2.7 \cdot 10^{-5}$ moles/microemulsion, while the total number of "undissociated salt species"/micelle was equal to the number of moles/microemulsion multiplied by Avogadro number and divided by the total number of micelles. Hence, for each individual ions we obtain: [$Ca^{2+}$] = 1.23-3.03; [$Na^+$] = 2.46-6.06; [$Cl^-$] = 2.46-6.06; [$CO_3^{2-}$] = 1.23-3.03 ions/micelle. The mixing of the individual microemulsions to start an experiment naturally halves these values. Although these are average values, one could actually expect a Poisson-like distribution of the number of ions across the water cores, since the micelles can exchange their contents with each other.

**Table S1. Summary of the structural parameters of the reverse micelles.**

| Polar phase: reverse micelle's core | Ionic surfactant: reverse micelle's shell | Apolar phase: oil medium |
|---|---|---|
| Water $\rho_{water}$ = 334 e$^-$/nm$^3$ | NaAOT In SAXS the shell includes only the polar heads. | Isooctane $\rho_{isooctane}$ = 241 e$^-$/nm$^3$ |
| The presence of dissolved salts at the concentrations used in our experiments increases $\rho_{water}$ by < 2% and is therefore neglected: | Polar hydrophilic head – essentially sulfonic group and sodium counter ion | |
| 0.15 mol/L CaCl$_2$, 1.012 g/cm$^3$: $\rho_{Ca,aq}$ = 337.4 e$^-$/nm$^3$ | $\rho_{head}$ = ~850 e$^-$/nm$^3$ $T_s$ = ~0.5 nm | |
| 0.15 mol/L Na$_2$CO$_3$, 1.015 g/cm$^3$: $\rho_{Na,aq}$ = 338.7 e$^-$/nm$^3$ | Apolar hydrophobic tail – essentially succinate group and 2-ethylhexoxy hydrocarbon chains $\rho_{tail}$ = ~250 e$^-$/nm$^3$ $L$ = 1.26 nm | |

**Table S2. Reference crystallographic information for calcite and several vaterite forms.**

| Phase | Structure origin | Space group | Database entry |
|---|---|---|---|
| calcite | Natural | R -3 c | AMCSD 0000098 |
| vaterite A | Synthetic | P 63/m m c | AMCSD 0009279 |
| vaterite B | Theoretical | P 65 2 2 | AMCSD 0004854 |
| vaterite C | Natural | P b n m | AMCSD 0019139 |
| vaterite D | Synthetic | C 1 2/c 1 | AMCSD 0019138 |
| vaterite E | Theoretical | P 65 | AMCSD 0019869 |
| vaterite F | Theoretical | P 32 2 1 | AMCSD 0019870 |

Fig. 1

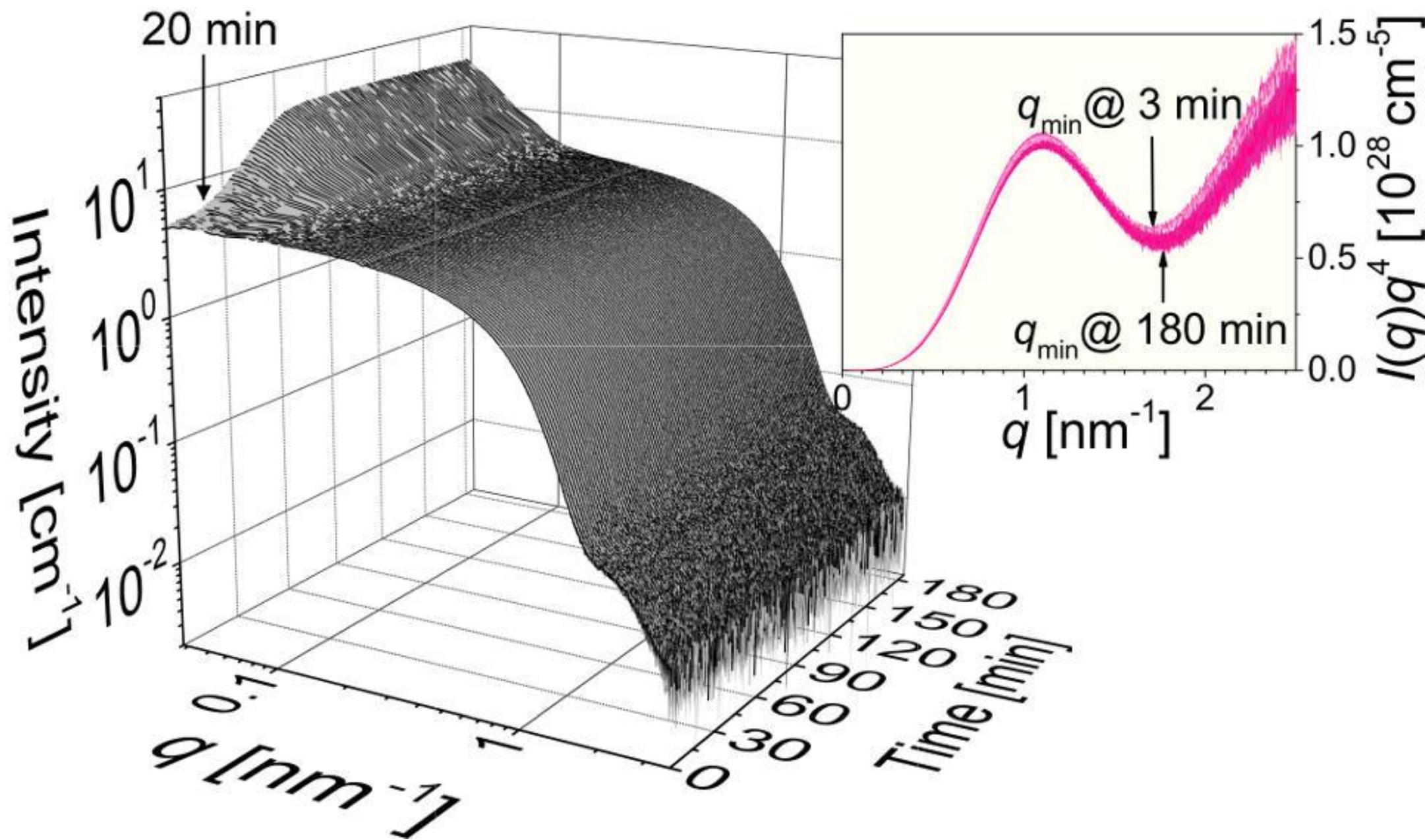

Fig. 2

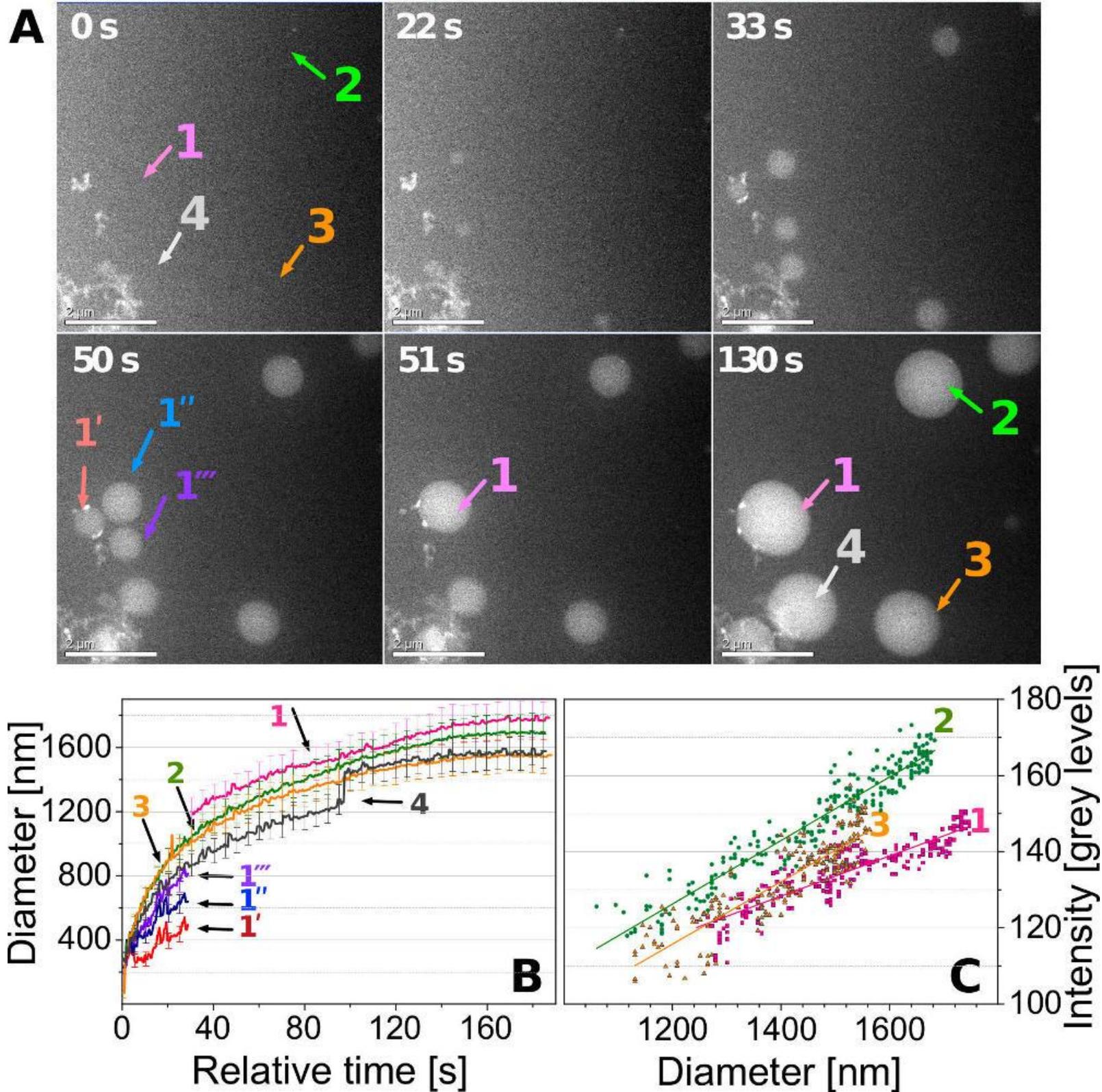

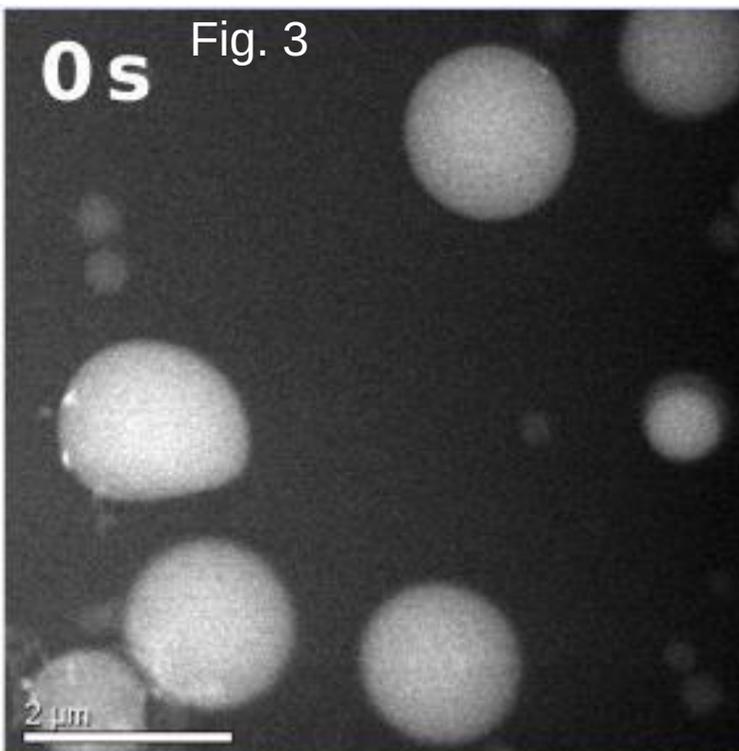 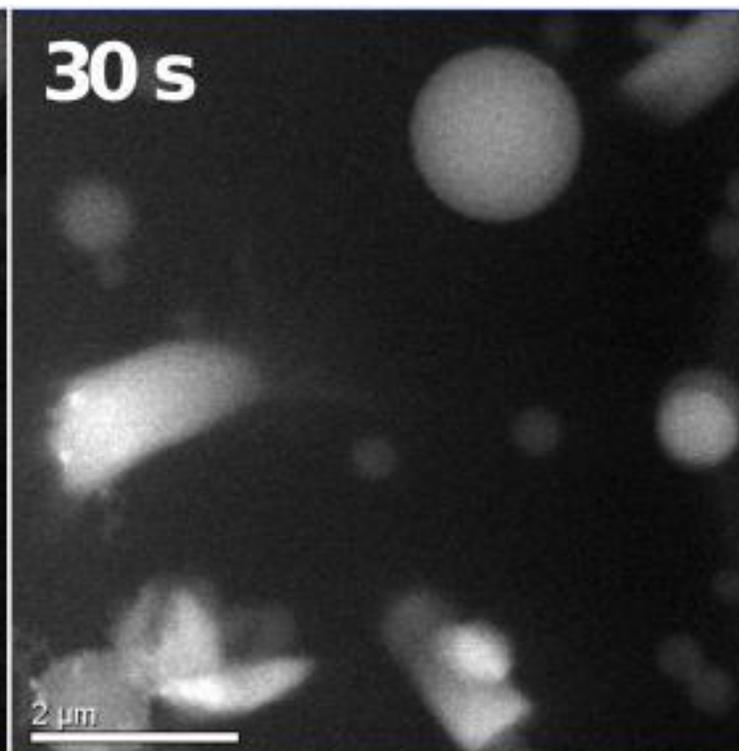 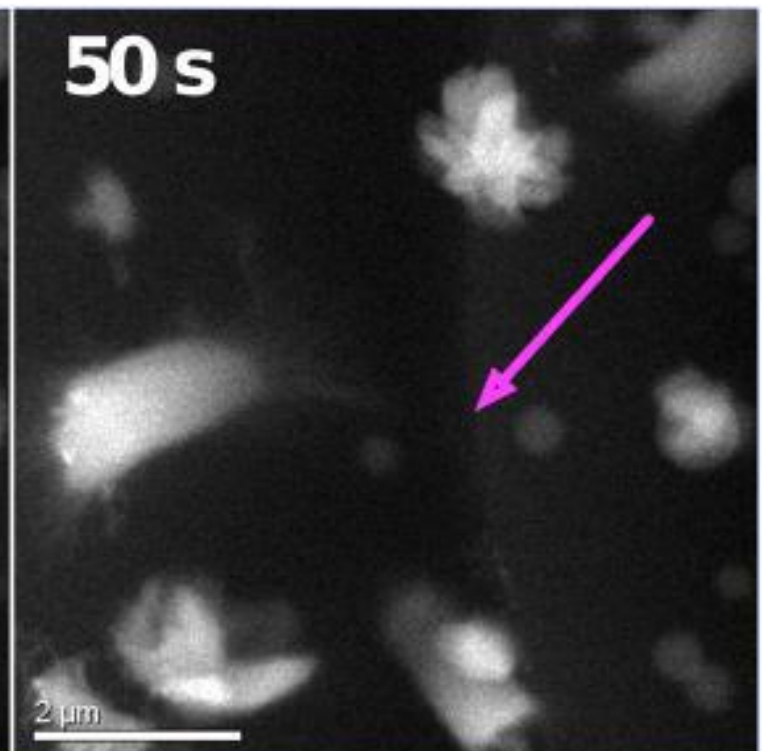
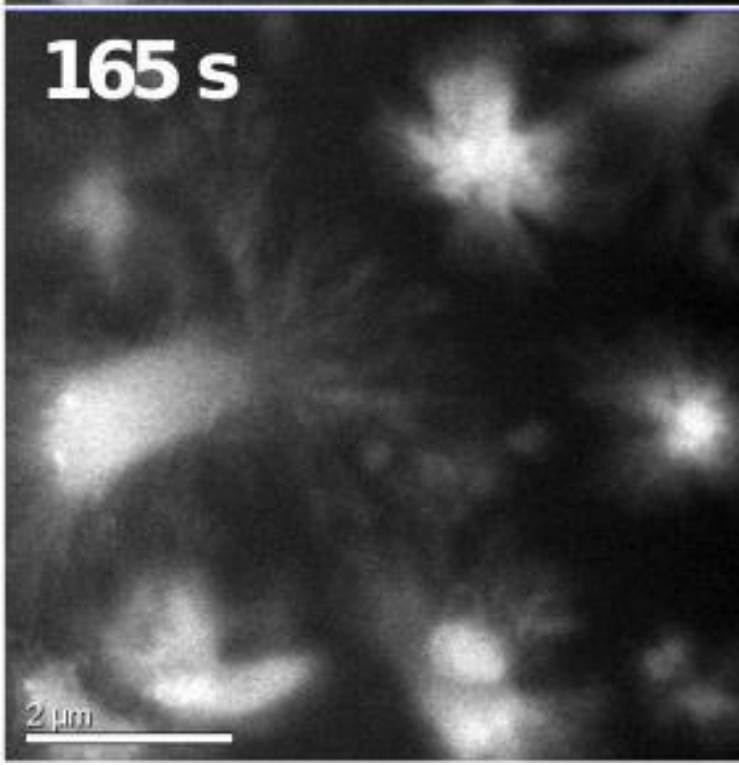 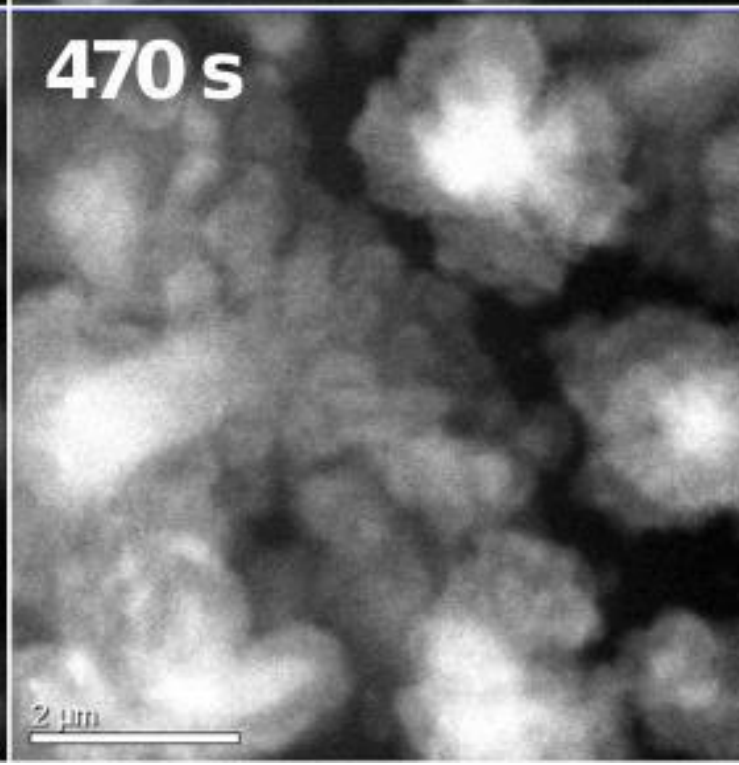 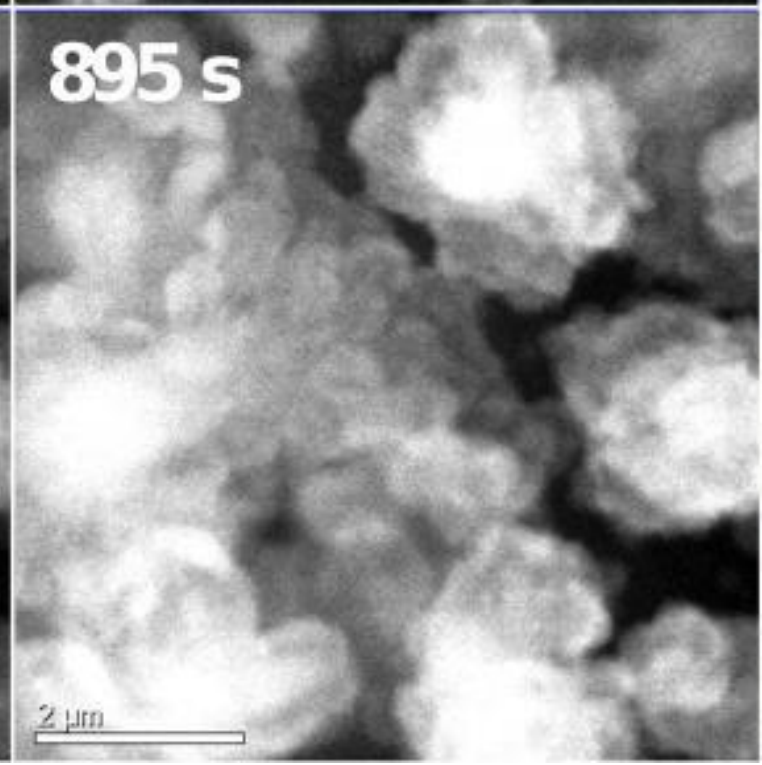

Fig. 3

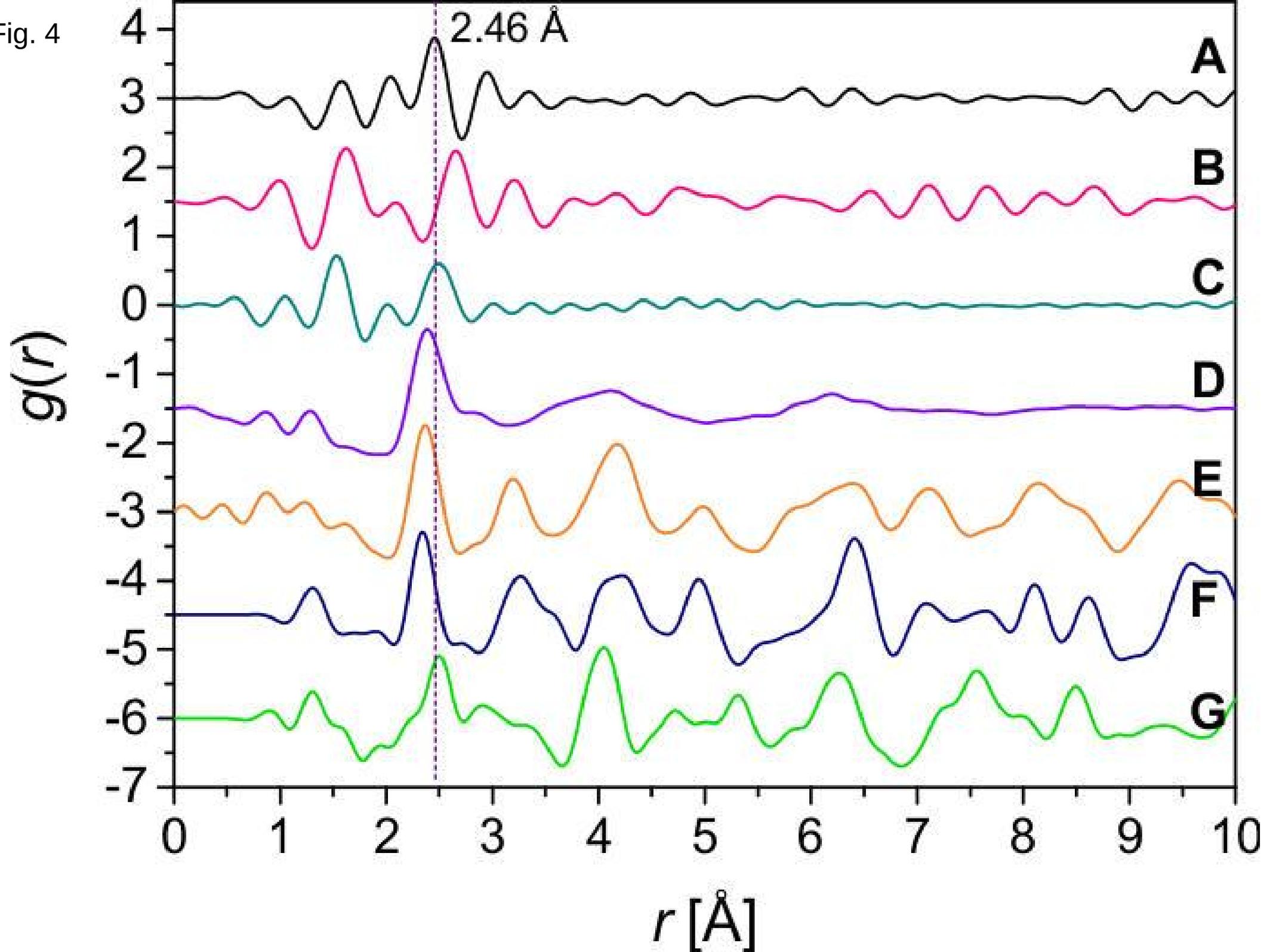

Fig. 5

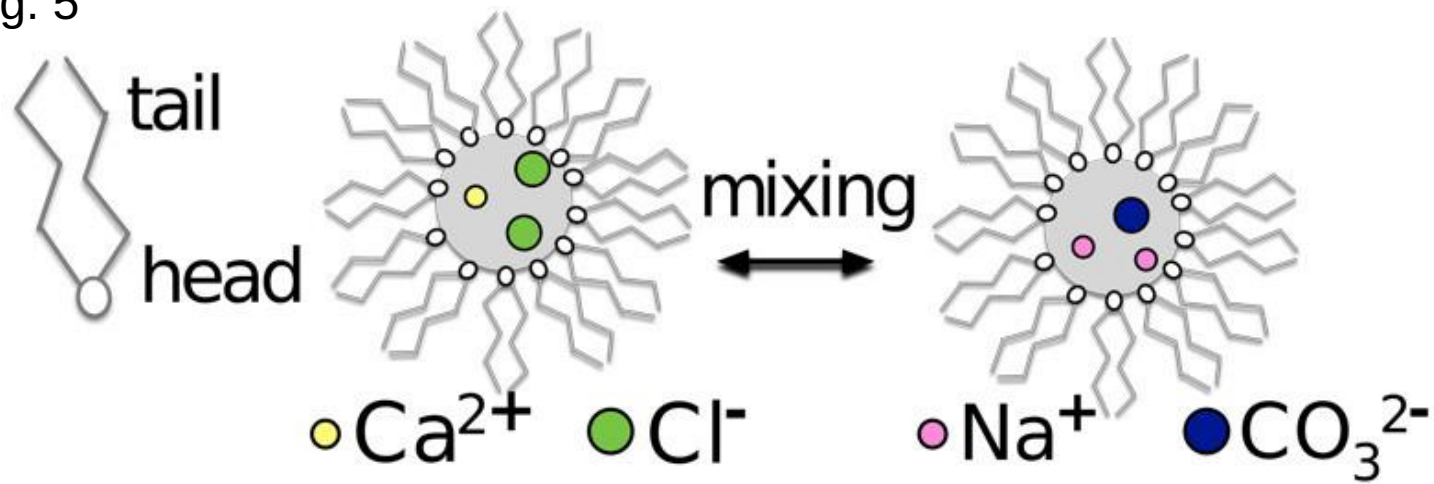
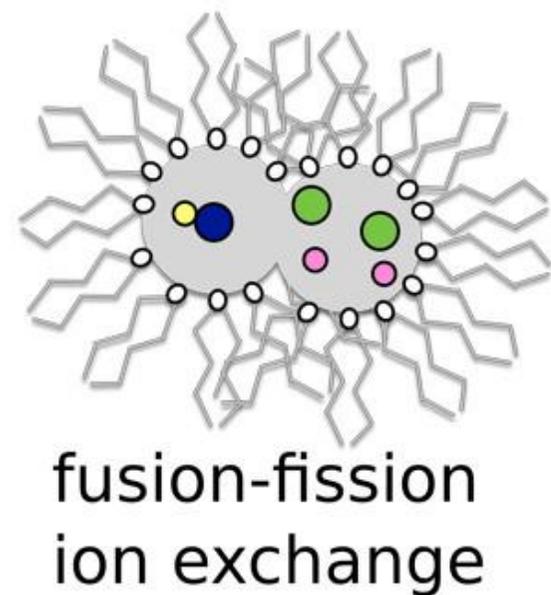
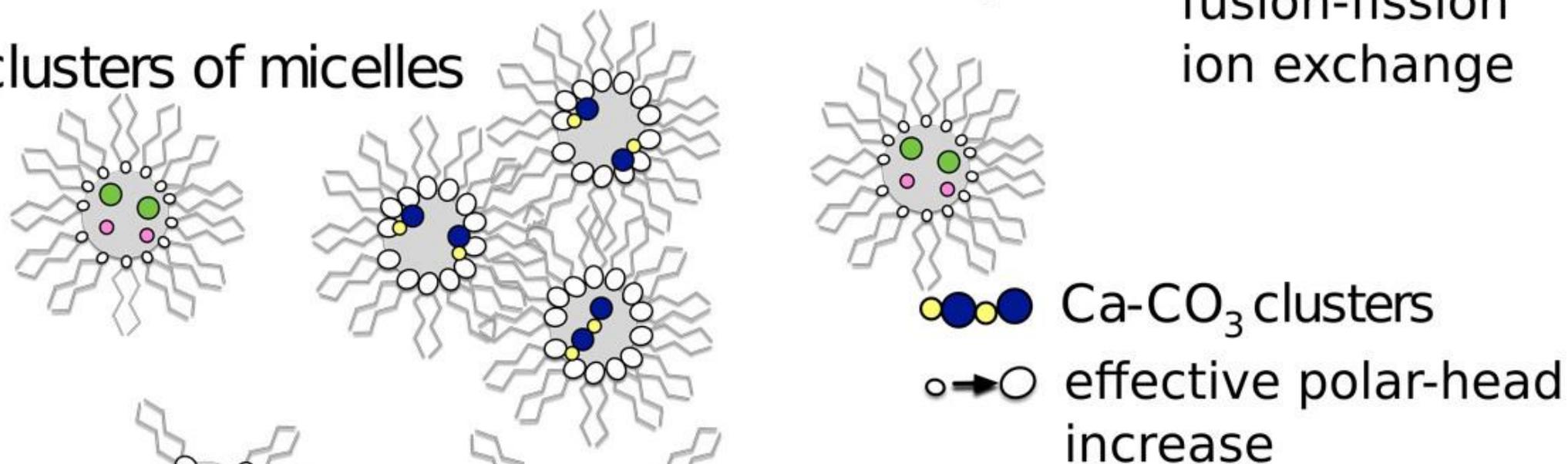
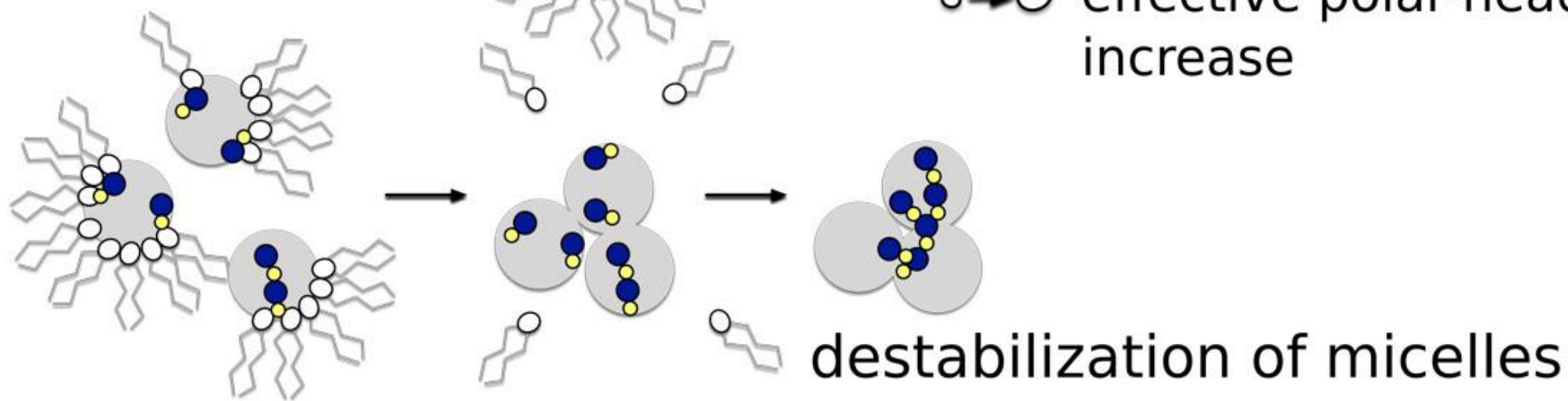

Fig. S1

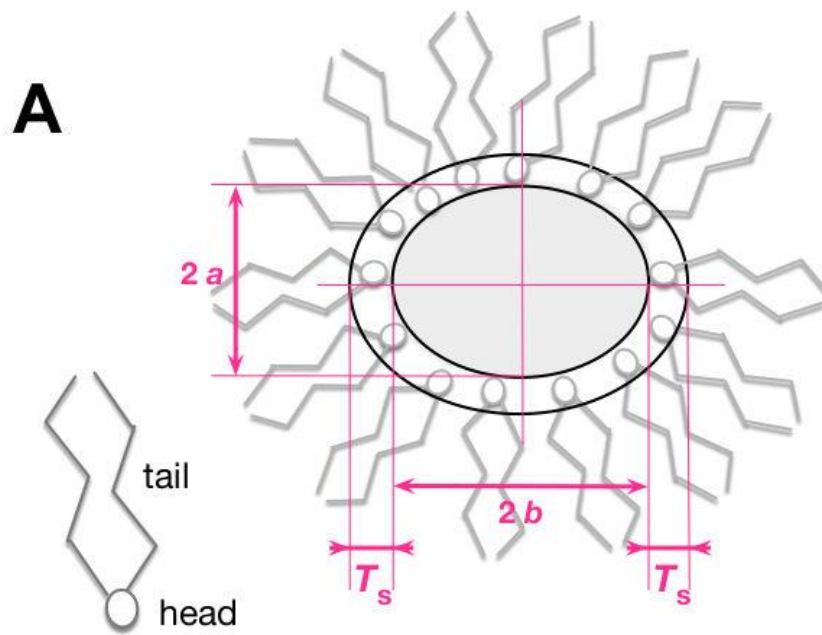
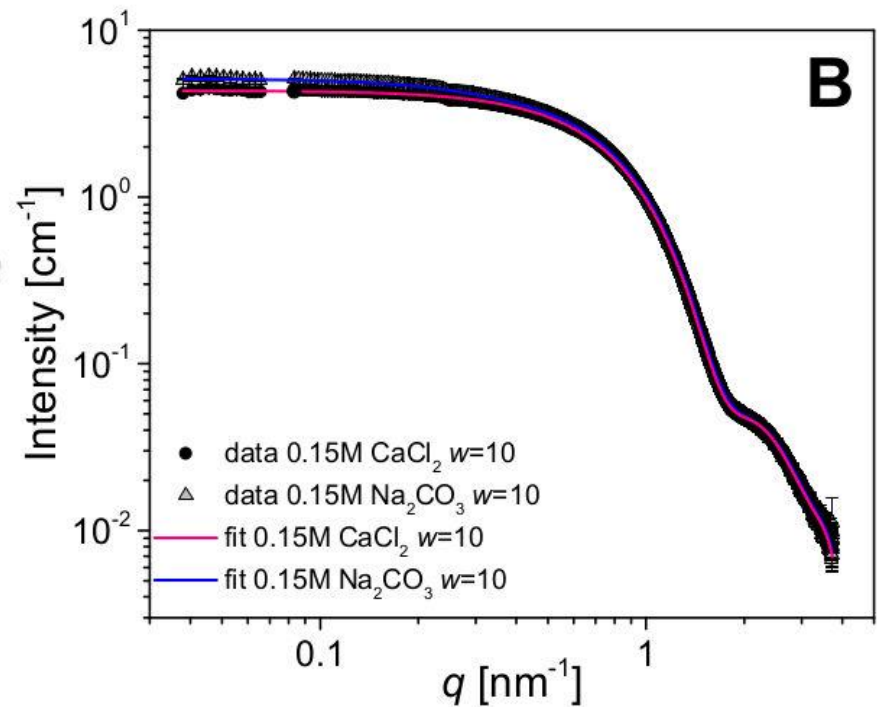
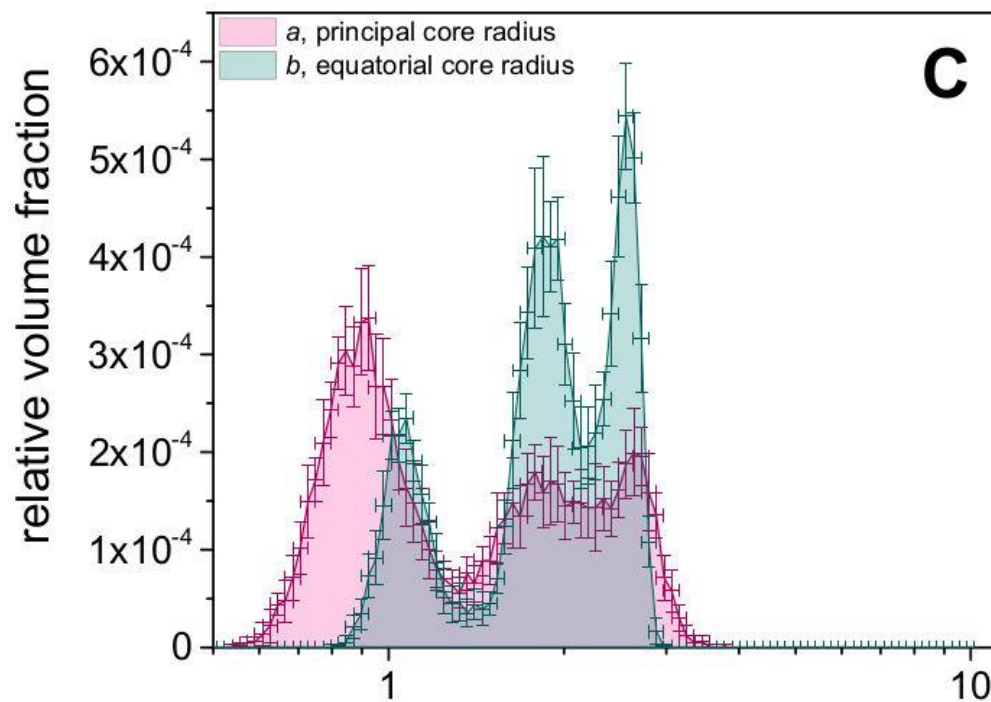
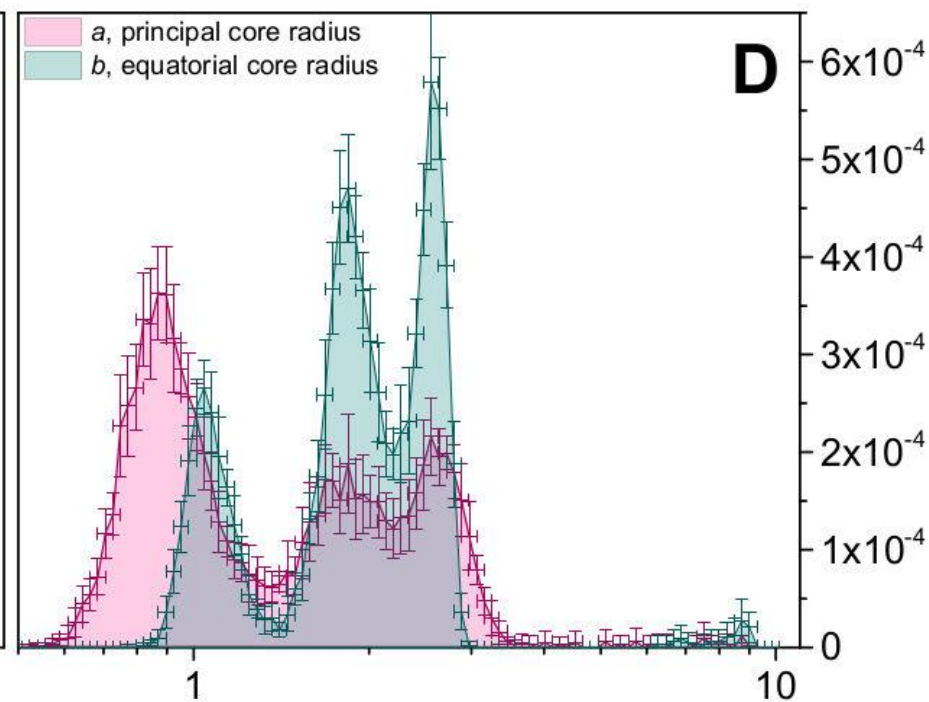

Fig. S2

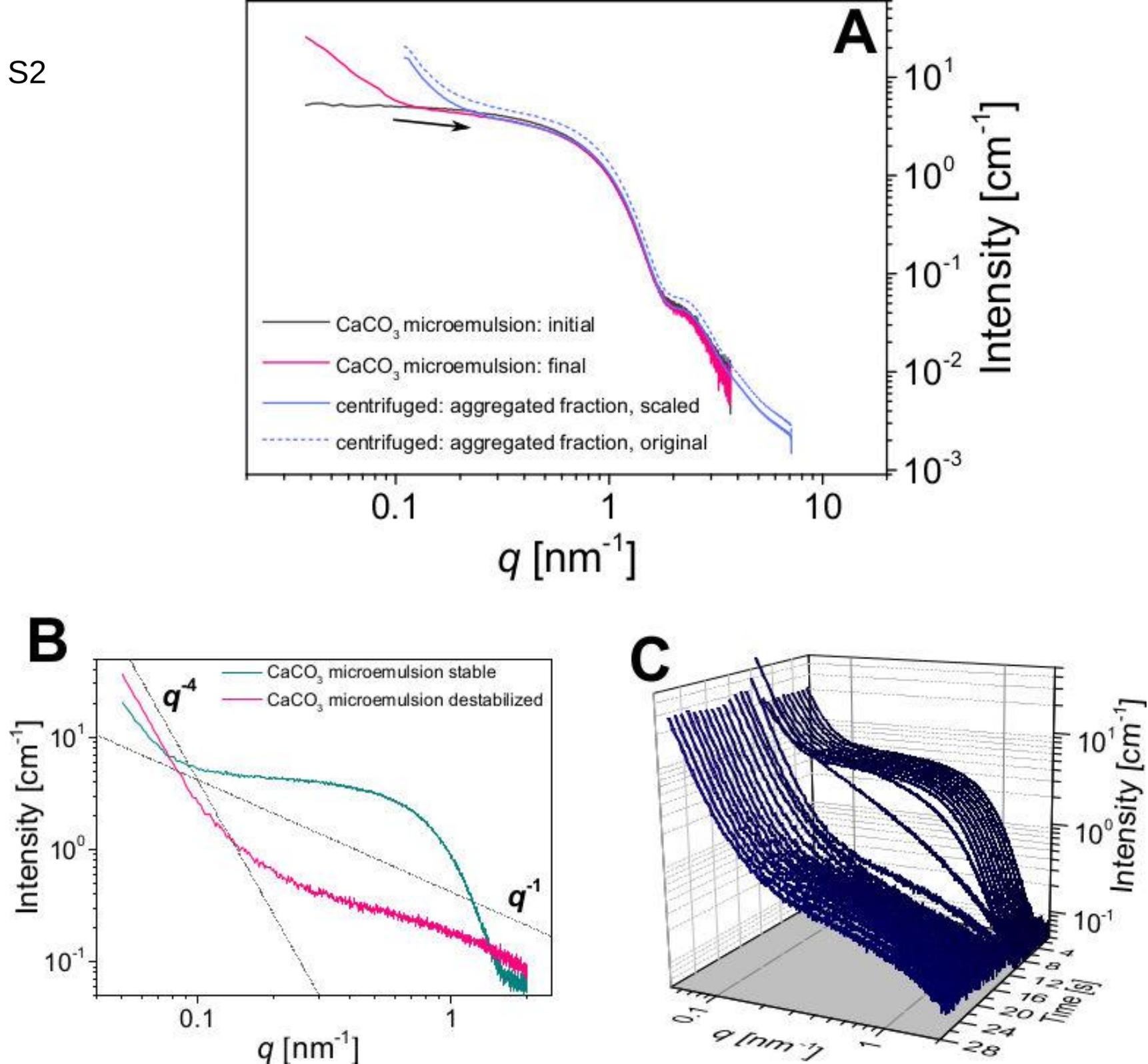

Fig. S3

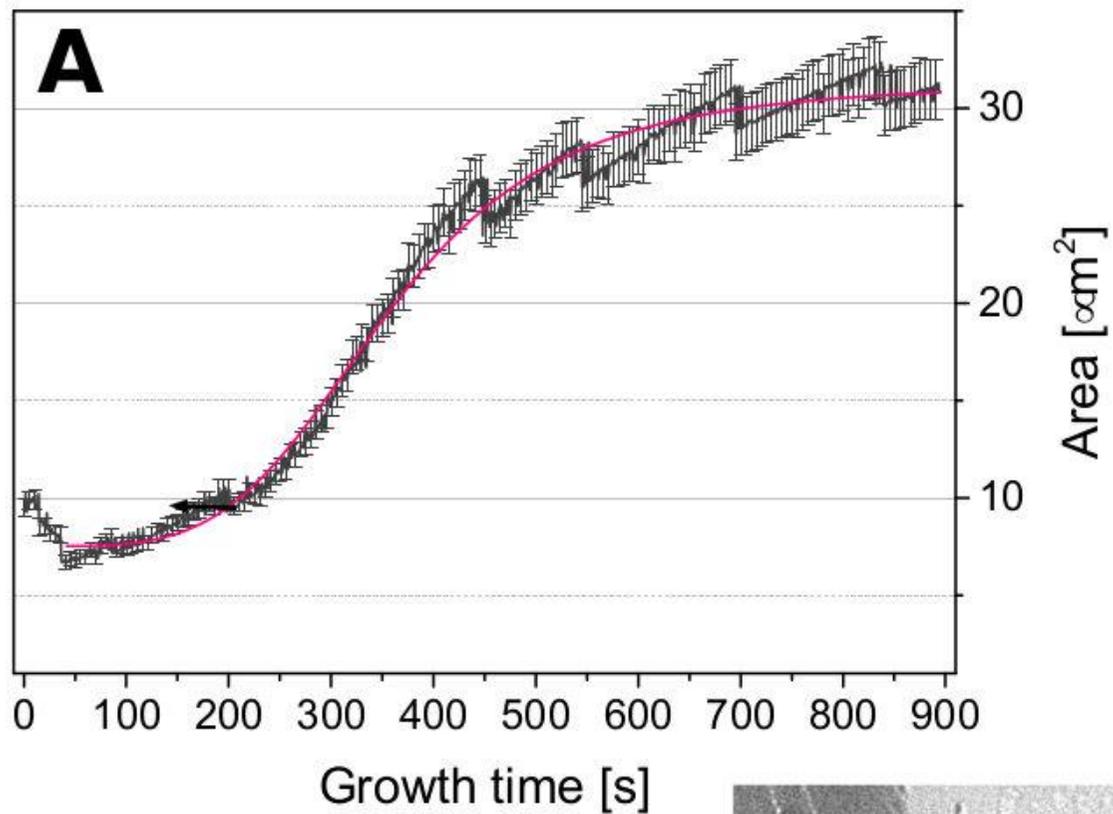
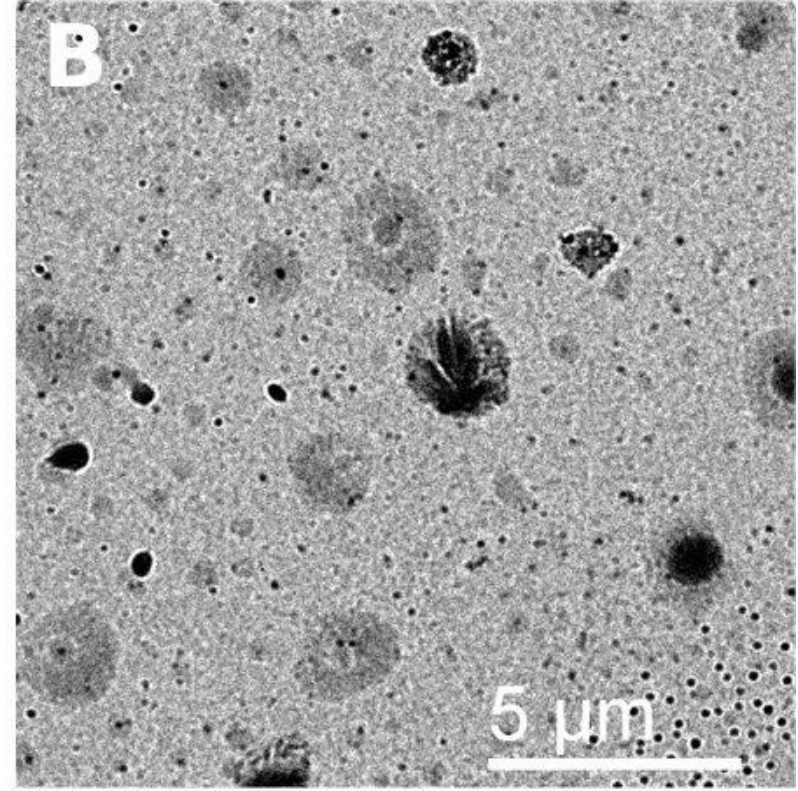
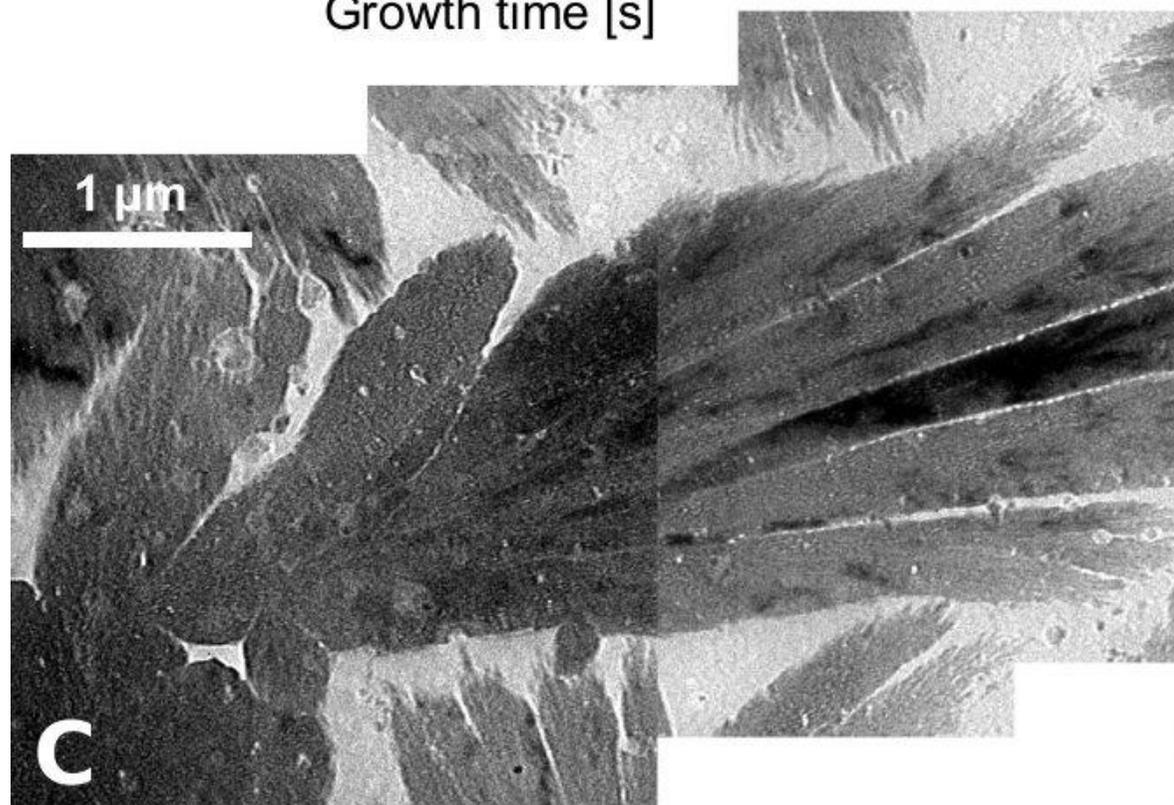
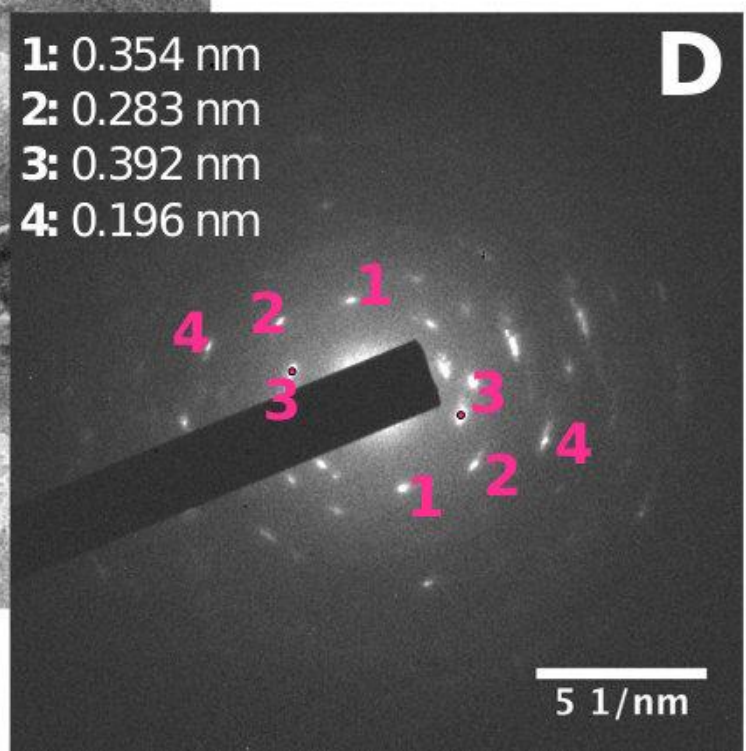

Fig. S4

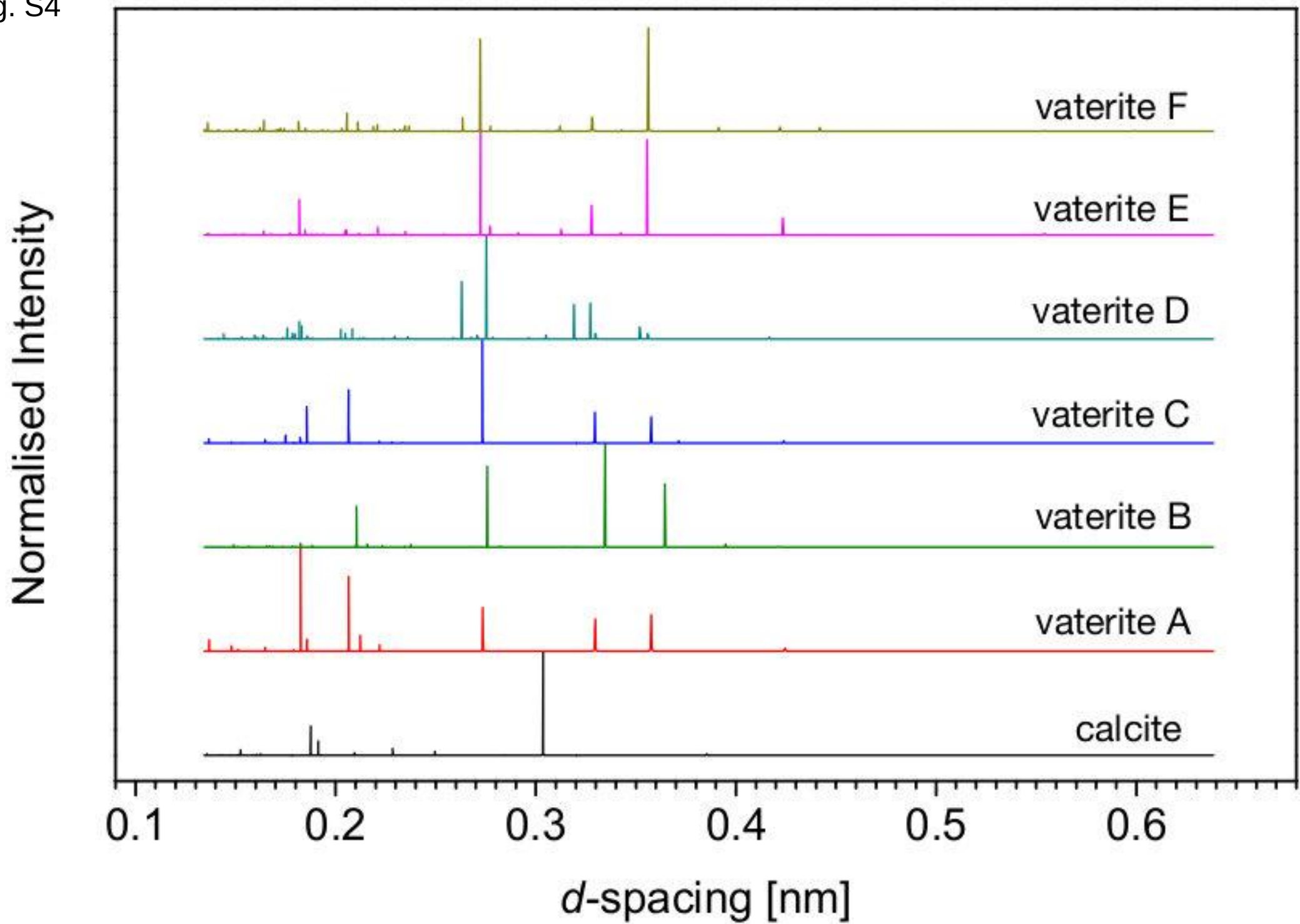

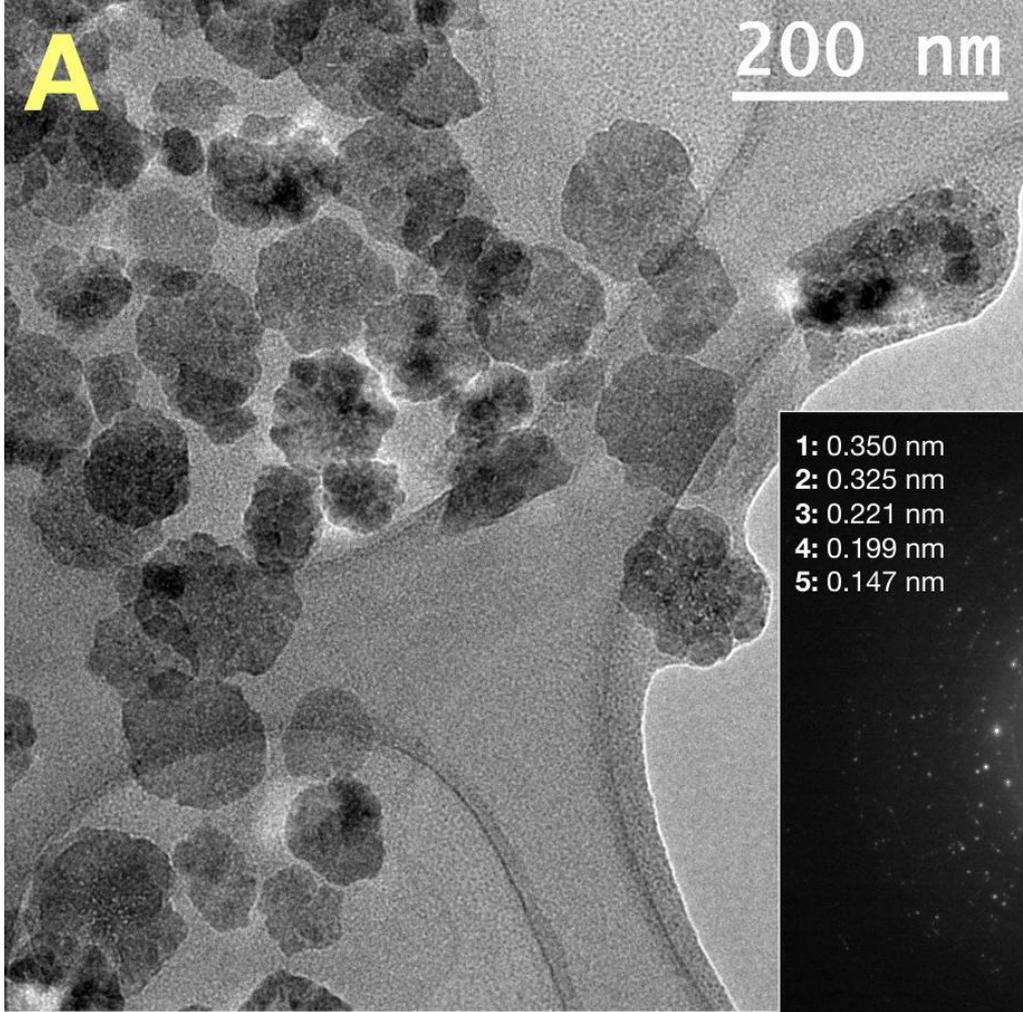
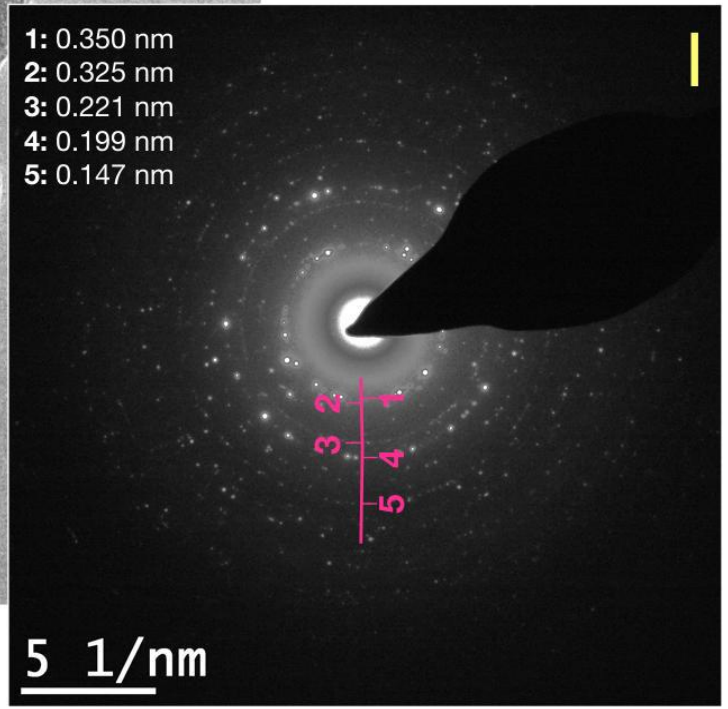
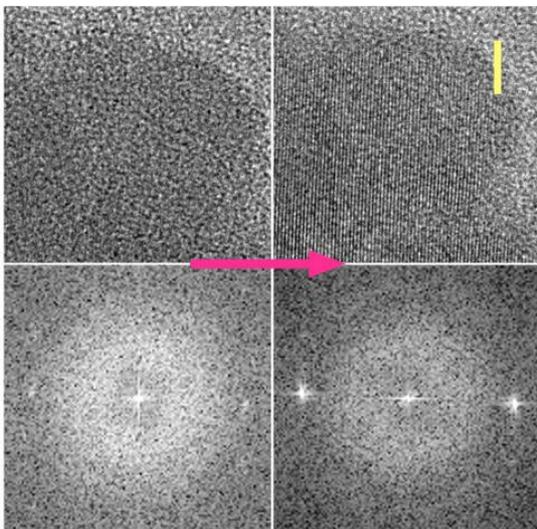
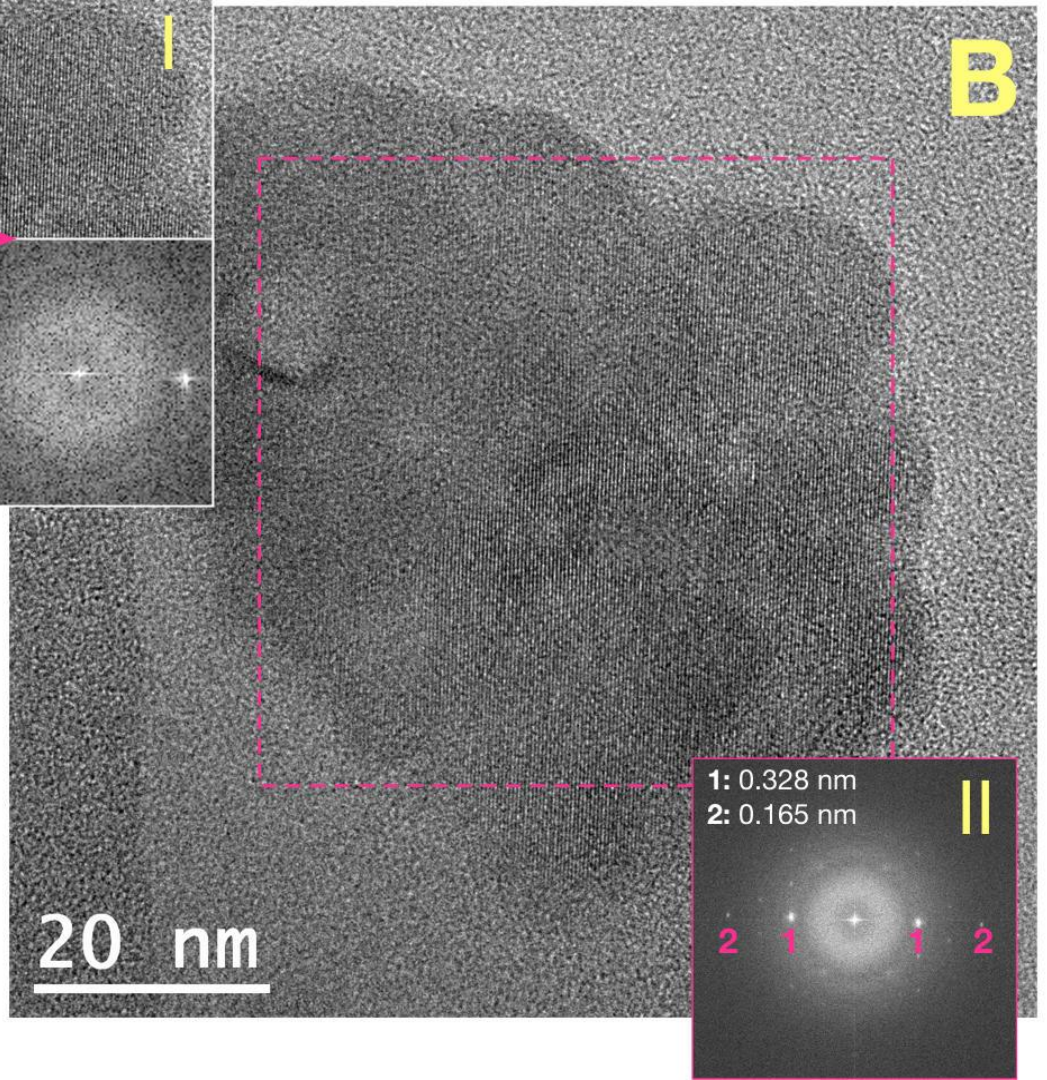

Fig. S6

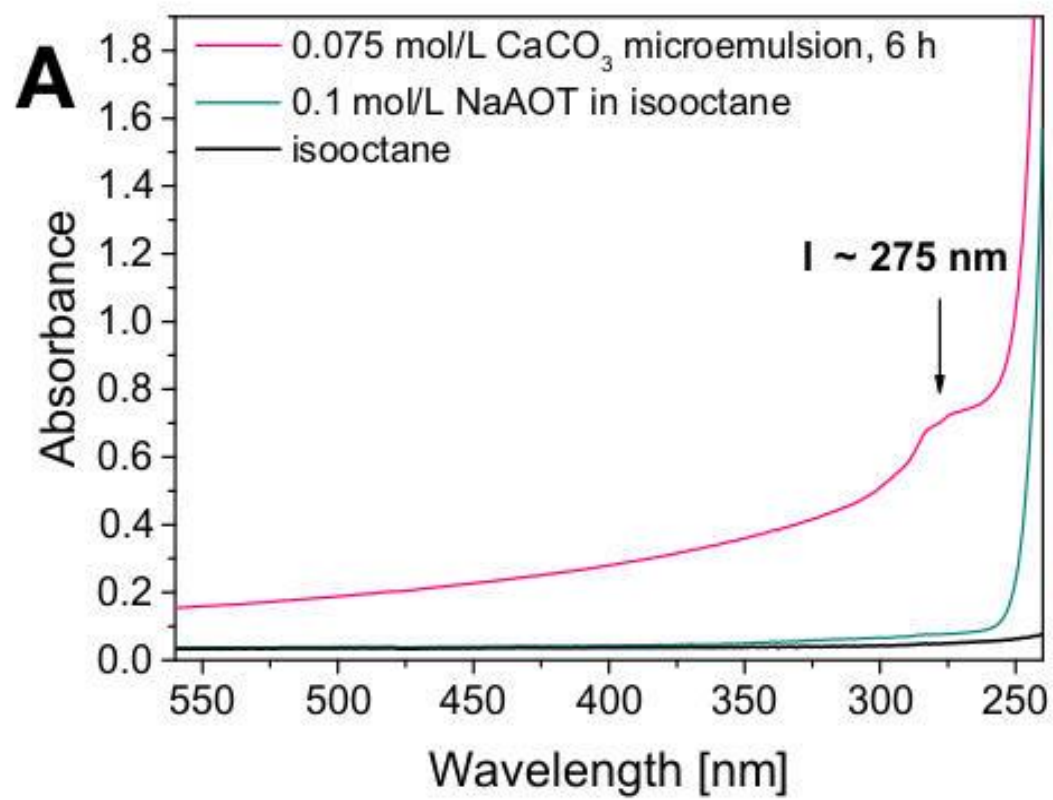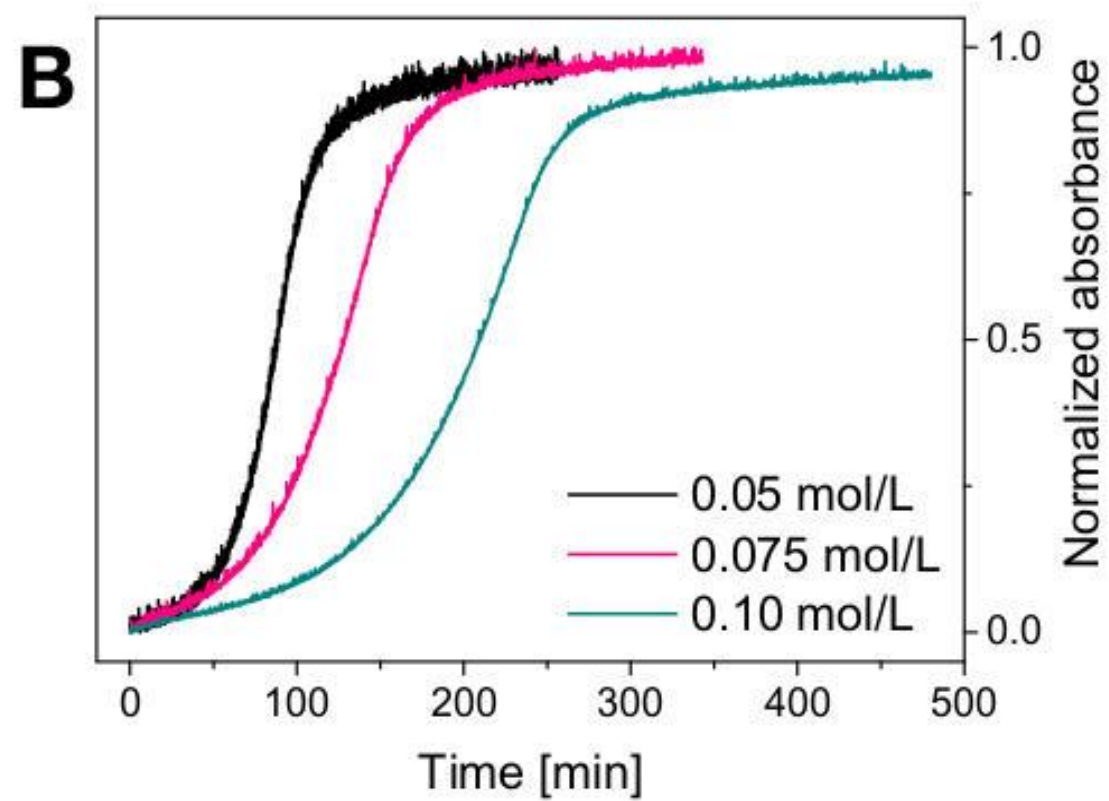